\documentclass[12pt]{elsarticle}

\usepackage{amssymb}
\usepackage{amsmath}
\usepackage{longtable}
\usepackage{booktabs}
\usepackage{multirow}
\usepackage{comment}
\usepackage{rotating}

\journal{}

\begin{document}

\begin{frontmatter}



\title{Algorithmic Energy Saving for Parallel Cholesky, LU, and QR Factorizations}


\author{Li Tan}\corref{cor_author}\author{Zizhong Chen}
\address{Department of Computer Science and Engineering, University of California, Riverside, CA, USA}
\cortext[cor_author]{Corresponding Author: darkwhite29@gmail.com}

\begin{abstract}
The pressing demands of improving energy efficiency for high performance scientific computing have motivated a large body of software-controlled hardware solutions using Dynamic Voltage and Frequency Scaling (DVFS) that strategically switch processors to low-power states, when the peak processor performance is not necessary. Although OS level solutions have demonstrated the effectiveness of saving energy in a black-box fashion, for applications with variable execution characteristics, the optimal energy efficiency can be blundered away due to defective prediction mechanism and untapped load imbalance. In this paper, we propose \texttt{TX}, a library level \emph{race-to-halt} DVFS scheduling approach that analyzes Task Dependency Set of each task in parallel Cholesky, LU, and QR factorizations to achieve substantial energy savings OS level solutions cannot fulfill. Partially giving up the generality of OS level solutions per requiring library level source modification, \texttt{TX} leverages algorithmic characteristics of the applications to gain greater energy savings. Experimental results on two power-aware clusters indicate that \texttt{TX} can save up to 17.8\% more energy than state-of-the-art OS level solutions with negligible 3.5\% on average performance loss.
\end{abstract}


\end{frontmatter}


\section{Introduction}

\subsection{Motivation}

With the growing prevalence of distributed-memory architectures, high performance scientific computing has been widely employed on supercomputers around the world ranked by the TOP500 list \cite{top500}. Considering the crucial fact that the costs of powering a supercomputer are rapidly increasing nowadays due to expansion of its size and duration in use, improving energy efficiency of high performance scientific applications has been regarded as a pressing issue to solve. The Green500 list \cite{green500}, ranks the top 500 supercomputers worldwide by performance-power ratio in six-month cycles. Consequently, root causes of high energy consumption while achieving performance efficiency in parallelism have been widely studied. With different focuses of studying, holistic hardware and software approaches for reducing energy costs of running high performance scientific applications have been extensively proposed. Software-controlled hardware solutions such as DVFS-directed (Dynamic Voltage and Frequency Scaling) energy efficient scheduling are deemed to be effective and lightweight \cite{ipdps05a} \cite{ppopp05} \cite{sc05a} \cite{sc05b} \cite{ppopp06a} \cite{sc07} \cite{ics09}. Performance and memory constraints have been considered as trade-offs for energy savings \cite{ccgrid07} \cite{icpp07} \cite{rtas09} \cite{ipccc13} \cite{iccs14}.

DVFS is a runtime technique that is able to switch operating frequency and supply voltage of a hardware component (CPU, GPU, memory, etc.) to different \emph{levels} (also known as \emph{gears} or \emph{operating points}) per workload characteristics of applications to gain energy savings dynamically. CPU and GPU are the most widely applied hardware components for saving energy via DVFS due to two primary reasons: (a) Compared to other components such as memory, CPU/GPU DVFS is easier to implement \cite{icac11} -- various handy DVFS APIs have been industrialized for CPU/GPU DVFS such as CPUFreq kernel infrastructure \cite{cpufreq} incorporated into the Linux kernel and NVIDIA System Management Interface (nvidia-smi) \cite{smi} for NVIDIA GPU; (b) CPU energy costs dominate the total system energy consumption \cite{tpds10} (CPU and GPU energy costs dominate if heterogeneous architectures are considered), and thus saving CPU and GPU energy greatly improves energy efficiency of the whole system. In this work, we focus on distributed-memory systems without GPU. For instance, energy saving opportunities can be exploited by reducing CPU frequency and voltage during non-CPU-bound operations such as large-message MPI communication, since generally execution time of such operations barely increases at a low-power state of CPU, in contrast to original runs at a high-power state. Given the fact that energy consumption equals product of average power consumption and execution time ($E = \overline{P} \times T$), and the assumption that dynamic power consumption of a CMOS-based processor is proportional to product of operating frequency and square of supply voltage ($P \propto fV^2$) \cite{ics02} \cite{sc05c}, energy savings can be effectively achieved using DVFS-directed strategical scheduling approaches with little performance loss.

Running on distributed-memory architectures, HPC applications can be organized and scheduled in the unit of task, a set of operations that are functionally executed as a whole. Different tasks within one process or across multiple processes may depend on each other due to \emph{intra-process} and \emph{inter-process} data dependencies. Parallelism of task-parallel algorithms and applications can be characterized by graph representations such as Directed Acyclic Graph (DAG), where data dependencies among parallel tasks are appropriately denoted by directed edges. DAG can be effective for analyzing parallelism present in HPC runs, which is greatly beneficial to achieving energy efficiency. As typical task-parallel algorithms for scientific computing, dense matrix factorizations in numerical linear algebra such as Cholesky, LU, and QR factorizations have been widely adopted to solve systems of linear equations. Empirically, as standard functionality, routines of dense matrix factorizations are provided by various software libraries of numerical linear algebra for distributed-memory multicore architectures such as ScaLAPACK \cite{scalapack}, DPLASMA \cite{dplasma}, and MAGMA \cite{magma}. Therefore, saving energy for parallel Cholesky, LU, and QR factorizations contributes significantly to the greenness of high performance scientific computing nowadays.

\subsection{Limitations of Existing Solutions}

Most existing energy saving solutions for high performance scientific applications are (combination of) variants of two classic approaches: (a) A Scheduled Communication (\emph{SC}) approach \cite{ipdps05a} \cite{cluster06} \cite{sc06} \cite{sc07} that keeps low CPU performance during communication and high CPU performance during computation, as large-message communication is not CPU-bound while computation is, and (b) a Critical Path (\emph{CP}) approach \cite{sc07} \cite{ics09} \cite{hpcs11} \cite{csrd12a} that guarantees that tasks on the CP run at the highest CPU frequency while reduces frequency \emph{appropriately} (i.e., without further delay to incur overall performance loss) for tasks off the CP to minimize slack.

Per the operating layer, existing solutions can be categorized into two types: OS level and application level. In general, OS level solutions feature two properties: (a) Working aside running applications and thus requiring no application-specific knowledge and no source modification, and (b) making online energy efficient scheduling decisions via dynamic monitoring and analysis. However, application level solutions statically utilize application-specific knowledge to perform specialized scheduling for saving energy, generally with source modification and recompilation (i.e., \emph{generality}) trade-offs. Although with high generality, OS level solutions may suffer from critical disadvantages as follows, and consequently are far from a sound and complete solution, for applications such as parallel Cholesky, LU, and QR factorizations in particular:

\vspace{1mm}
\noindent\textsc{\textbf{Effectiveness}}. Although intended to be effective for general applications, OS level approaches rely heavily on the underlying workload prediction mechanism, due to lack of knowledge of application characteristics. One prediction mechanism can work well for a specific type of applications sharing similar characteristics, but can be error-prone for other applications, in particular, applications with random/variable execution characteristics where the prediction mechanism performs poorly. Algorithms presented in \cite{sc06} \cite{sc07} \cite{ics09} predict execution characteristics of the upcoming interval (i.e., a fixed time slice) according to recent intervals. This prediction mechanism is based on a simple assumption that task behavior is identical every time a task is executed \cite{ics09}. However, it can be defective for applications with random/variable execution characteristics, such as matrix factorizations, where the remaining unfinished matrices become smaller as the factorizations proceed. In other words, length variation of iterations of the core loop for matrix factorizations can make the above prediction mechanism inaccurate, which invalidates potential energy savings. Moreover, since new technologies such as Hyper-Threading \cite{ht} have emerged for exploiting thread level parallelism on distributed-memory multicore systems, e.g., MPI programs can be parallelized on local multicore processors using OpenMP, behavior among parallel tasks can vary due to more non-deterministic events occurred in the multithreaded environment.

Further, the OS level prediction can be costly and thus energy savings are diminished. Recall that OS level solutions must predict execution details in the next interval using prior execution information. However, execution history in some cases may not necessarily be a reliable source for workload prediction, e.g., for applications with fluctuating runtime patterns at the beginning of the execution. As such, it can be time-consuming for obtaining accurate prediction results. Since during the prediction phase no energy savings can be fulfilled, considerable potential energy savings can be wasted for accurate but lengthy prediction.

\vspace{1mm}
\noindent\textsc{\textbf{Completeness}}. OS level solutions only work when tasks such as computation and communication are being executed, energy saving opportunities are untapped during the time otherwise. Empirically, even though load balancing techniques have been leveraged, due to data dependencies among tasks and load imbalance that is not completely eliminated, not all tasks in different processes across nodes can start to work and finish at the same time. More energy can be saved for tasks waiting at the beginning of an execution, and for the last task of one process finishing earlier than that of other processes across nodes (details are illustrated in Figure \ref{Cholesky_DAG}). Restricted by the daemon-based nature of working aside real running tasks, OS level solutions cannot attain energy savings for such tasks.

\subsection{Our Contributions}

In this paper, we propose a library level \emph{race-to-halt} DVFS scheduling approach via Task Dependency Set (TDS) analysis based on algorithmic characteristics, namely \texttt{TX} (TDS-based race-to-halt), to save energy for task-parallel applications running on distributed-memory architectures, taking parallel Cholesky, LU, and QR factorizations for example. The idea of library level \emph{race-to-halt} scheduling is intended for any task-parallel programming models where data flow analysis can be applied. The use of TDS analysis as a compiler technique allows possible extension of this work to a general compiler-based approach based on static analysis only. In summary, the contributions of this paper are as follows:

\begin{itemize}
\item Compared to application level solutions, for widely used software libraries such as numerical linear algebra libraries, \texttt{TX} restricts source modification and recompilation at library level, and replacement of the energy efficient version of the libraries is allowed at link time (i.e., with \emph{partial loss of generality}). No further source modification and recompilation are needed for applications where the libraries are called;
\item Compared to OS level solutions, \texttt{TX} is able to achieve substantial energy savings for parallel Cholesky, LU, and QR factorizations (i.e., with \emph{higher energy efficiency}), since via algorithmic TDS analysis, \texttt{TX} circumvents the defective prediction mechanism at OS level, and also manages to save more energy from possible load imbalance;
\item We formally model that \texttt{TX} is comparable to the \emph{CP} approach in energy saving capability under the circumstance of current CMOS technologies that allow insignificant variation of supply voltage as operating frequency scales via DVFS;
\item With negligible performance loss (3.5\% on average), on two power-aware clusters, \texttt{TX} is evaluated to achieve up to 33.8\% energy savings compared to original runs of different-scale matrix factorizations, and save up to 17.8\% and 15.9\% more energy than state-of-the-art OS level \emph{SC} and \emph{CP} approaches, respectively.
\end{itemize}

The rest of this paper is organized as follows. Section 2 introduces basics of parallel Cholesky, LU, and QR factorizations. We present TDS and CP in section 3, and our \texttt{TX} approach in section 4. Implementation details and experimental results are provided in section 5. Section 6 discusses related work and section 7 concludes.

\section{Background: Parallel Cholesky, LU, and QR Factorizations}

As classic dense numerical linear algebra operations for solving systems of linear equations, such as $Ax = b$ where $A$ is a given coefficient matrix and $b$ is a given vector, Cholesky factorization applies to the case that $A$ is a symmetric positive definite matrix, while LU and QR factorizations apply to any general $M \times N$ matrices. The goal of these operations is to factorize $A$ into the form $LL^T$ where $L$ is lower triangular and $L^T$ is the transpose of $L$, the form $LU$ where $L$ is unit lower triangular and $U$ is upper triangular, and the form $QR$ where $Q$ is orthogonal and $R$ is upper triangular, respectively. Thus from $LL^Tx = b$, $LUx = b$, $QRx = b$, $x$ can be easily solved via forward substitution and back substitution. In practice, parallel Cholesky, LU, and QR factorizations are widely employed in extensive areas of high performance scientific computing. Various software libraries of numerical linear algebra for distributed multicore scientific computing such as ScaLAPACK \cite{scalapack}, DPLASMA \cite{dplasma}, and MAGMA \cite{magma} provide routines of these matrix factorizations as standard functionality. In this section, we present basics of parallel Cholesky, LU, and QR factorizations, and introduce an effective graph representation for parallelism present during the execution of these matrix factorizations.

\subsection{2-D Block Cyclic Data Distribution}

Regardless of the fact that nowadays matrix involved in many parallel numerical linear algebra operations is too large to fit into the memory of single node, the essence of a parallel algorithm is how it partitions workloads into a cluster of computing nodes as balanced as possible to better exploit parallelism, which is also referred to as load balancing. In numerical linear algebra operations, distributing a global matrix into a process grid in a linear fashion does not benefit a lot from parallelism, since although the data allocated into each computing node are balanced in terms of amount, parallel execution of computation and communication are restricted by frequently arising data dependencies among tasks performed by different processes on different nodes.

\begin{figure}[h]
\centering
\begin{tabular}{cc}
\includegraphics[width=1.524in]{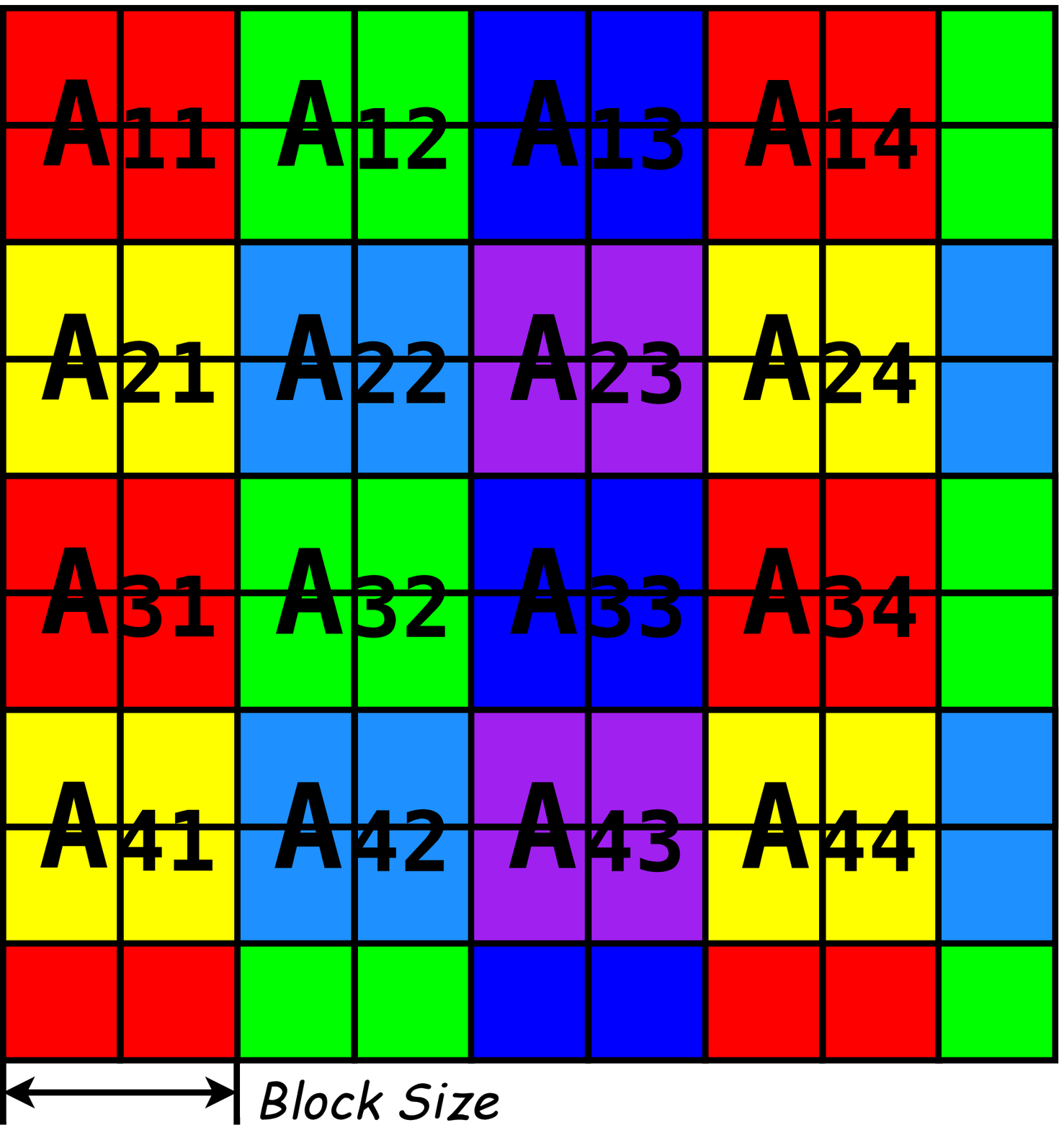} & \includegraphics[width=1.5in]{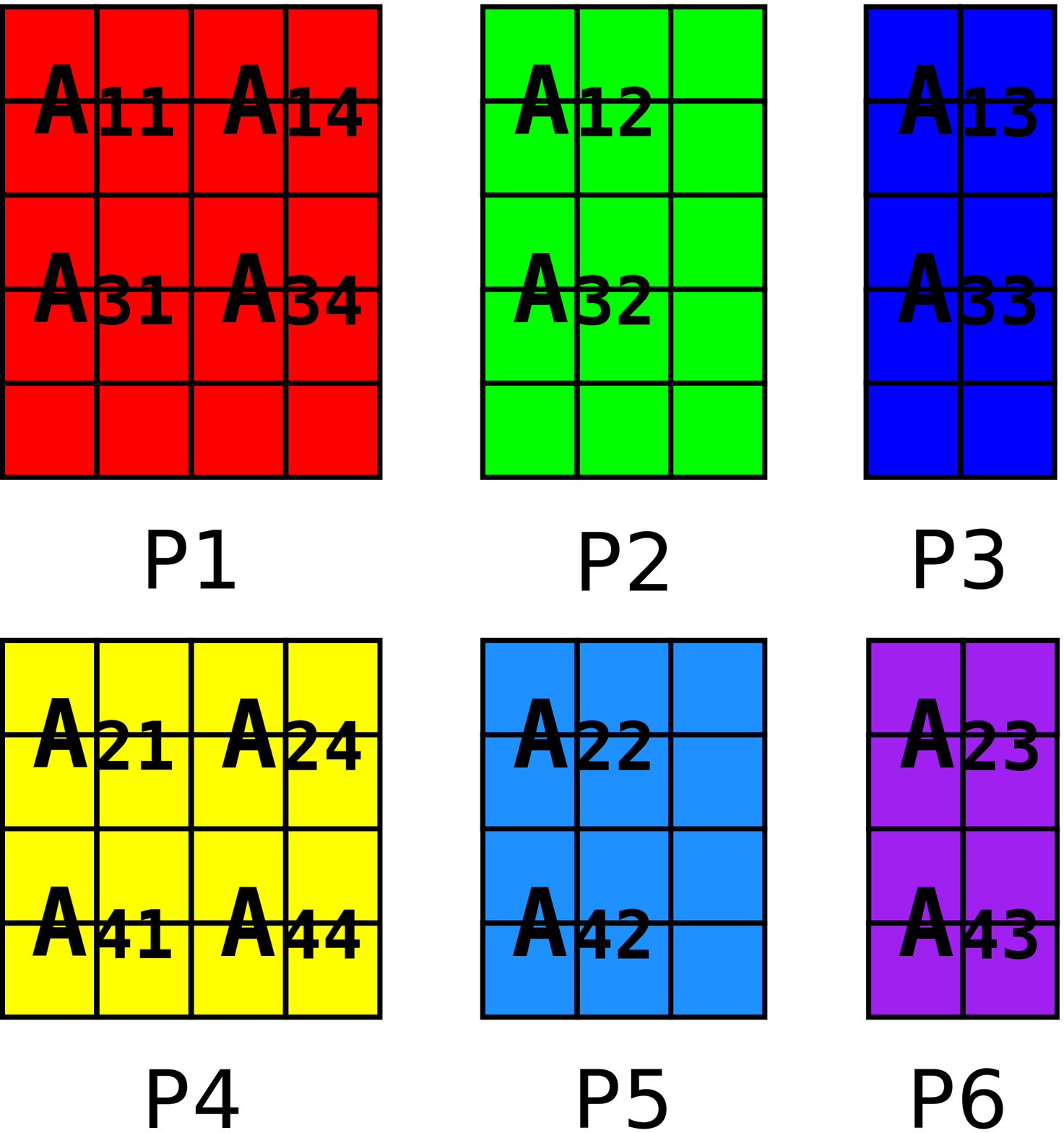}\\
\footnotesize(a) Global View\normalsize & \footnotesize(b) Local View\normalsize\\
\end{tabular}
\caption{2-D Block Cyclic Data Distribution on a 2 $\times$ 3 Process Grid in Global View and Local View.}
\label{2-D_block_cyclic_data_distribution}
\end{figure}

As shown in Figure \ref{2-D_block_cyclic_data_distribution}, \emph{2-D block cyclic data distribution}, an effective way for load balancing, has been widely used in various numerical linear algebra libraries, such as HPL \cite{hpl}, ScaLAPACK \cite{scalapack}, DPLASMA \cite{dplasma}, and MAGMA \cite{magma}. Specifically, the global matrix is partitioned in a two-dimensional block cyclic fashion, and blocks are mapped into different nodes cyclically along both rows and columns of the process grid, so that tasks without data dependencies are able to be executed by multiple nodes simultaneously to achieve parallelism. Figure \ref{2-D_block_cyclic_data_distribution} (a) shows the partitioned global matrix in a global view, while Figure \ref{2-D_block_cyclic_data_distribution} (b) depicts local matrices residing in different nodes individually, where the global matrix elements on one node are accessed periodically throughout the execution to balance the workloads in different nodes, instead of all at once as in the linear data distribution. The granularity of partitioning is determined by a \emph{block size}, either specified by the user, or automatically tuned according to hardware configuration or set by default by the application itself. In practice, the global matrix size may not necessarily be multiples of the chosen block size. Typically, due to uneven division, the existence of remainder matrices, i.e., the last row and the last column in Figure \ref{2-D_block_cyclic_data_distribution} (a), barely affects parallelism empirically given a fine-grained partition.

\subsection{DAG Representation of Parallel Cholesky, LU, and QR Factorizations}

A well-designed partitioning and highly-efficient parallel algorithms of computation and communication substantially determine energy and performance efficiency of task-parallel applications. For such purposes, classic implementations of parallel Cholesky, LU, and QR factorizations are as follows: (a) Partition the global matrix into a cluster of computing nodes as a process grid using 2-D block cyclic data distribution \cite{ics11} for load balancing; (b) perform local \emph{diagonal matrix} factorizations in each node individually and communicate factorized local matrices to other nodes for \emph{panel matrix} solving and \emph{trailing matrix} updating, as shown in Figure \ref{stepwise_LU_factorization}, a stepwise LU factorization without pivoting. Due to frequently-arising data dependencies, parallel execution of parallel Cholesky, LU, and QR factorizations can be characterized using Directed Acyclic Graph (DAG), where data dependencies among parallel tasks are appropriately represented. DAG for parallel Cholesky, LU, and QR factorizations is formally defined below:

\begin{figure}[h]
\centering
\includegraphics[width=5.29in]{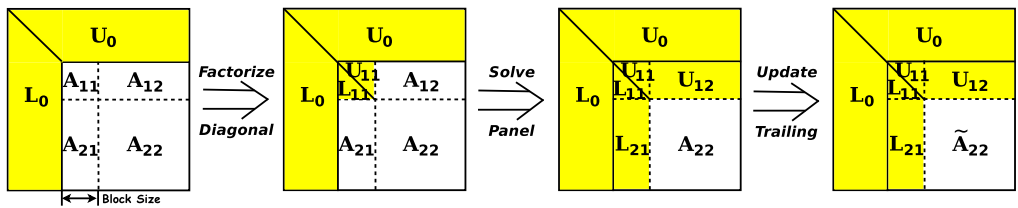}\\
\caption{Stepwise Illustration of LU Factorization without Pivoting.}
\label{stepwise_LU_factorization}
\end{figure}

\vspace{1mm}
\noindent\textsc{\textbf{Definition 1}}. Data dependencies among parallel tasks of parallel Cholesky, LU, and QR factorizations running on a distributed-memory computing system are modeled by a Directed Acyclic Graph (DAG) $G=(V,E)$, where each node $v \in V$ denotes a task of Cholesky/LU/QR factorization, and each directed edge $e \in E$ represents a dynamic data dependency from task $t_j$ to task $t_i$ that both tasks manipulate on either different \emph{intra-process} or \emph{inter-process} local matrices (i.e., an \emph{explicit} dependency) or the same \emph{intra-process} local matrix (i.e., an \emph{implicit} dependency), denoted by $t_i \rightarrow t_j$.

\begin{figure*}
\scriptsize
\begin{align*}
\hspace{-5.6em}
\begin{pmatrix}
A_{11} & A_{21}^T & A_{31}^T & A_{41}^T\\
A_{21} & A_{22}   & A_{32}^T & A_{42}^T\\
A_{31} & A_{32}   & A_{33}   & A_{43}^T\\
A_{41} & A_{42}   & A_{43}   & A_{44}
\end{pmatrix}
&=
\begin{pmatrix}
L_{11} & 0        & 0        & 0\\
L_{21} & L_{22}   & 0        & 0\\
L_{31} & L_{32}   & L_{33}   & 0\\
L_{41} & L_{42}   & L_{43}   & L_{44}
\end{pmatrix}
\times
\begin{pmatrix}
L_{11}^T & L_{21}^T & L_{31}^T & L_{41}^T\\
0        & L_{22}^T & L_{32}^T & L_{42}^T\\
0        & 0        & L_{33}^T & L_{43}^T\\
0        & 0        & 0        & L_{44}^T
\end{pmatrix}
\\
&=
\begin{pmatrix}
L_{11}L_{11}^T & L_{11}L_{21}^T                & L_{11}L_{31}^T                               & L_{11}L_{41}^T\\
L_{21}L_{11}^T & L_{21}L_{21}^T+L_{22}L_{22}^T & L_{21}L_{31}^T+L_{22}L_{32}^T                & L_{21}L_{41}^T+L_{22}L_{42}^T\\
L_{31}L_{11}^T & L_{31}L_{21}^T+L_{32}L_{22}^T & L_{31}L_{31}^T+L_{32}L_{32}^T+L_{33}L_{33}^T & L_{31}L_{41}^T+L_{32}L_{42}^T+L_{33}L_{43}^T\\
L_{41}L_{11}^T & L_{41}L_{21}^T+L_{42}L_{22}^T & L_{41}L_{31}^T+L_{42}L_{32}^T+L_{43}L_{33}^T & L_{41}L_{41}^T+L_{42}L_{42}^T+L_{43}L_{43}^T+L_{44}L_{44}^T
\end{pmatrix}
\end{align*}
\caption{Matrix Representation of a 4 $\times$ 4 Blocked Cholesky Factorization (We henceforth take parallel Cholesky factorization for example due to algorithmic similarity among three types of matrix factorizations).}
\label{Cholesky_matrix}
\end{figure*}

\vspace{1mm}
\noindent\textsc{\textbf{Example}}. Due to similarity among the three matrix factorizations and space limitation, we henceforth take parallel Cholesky factorization for example to elaborate our approach. Consider a 4 $\times$ 4 blocked Cholesky factorization as given in Figure \ref{Cholesky_matrix}. The outcome of the task factorizing $A_{11}$, i.e., $L_{11}$, is used in the tasks solving local matrices $L_{21}$, $L_{31}$, and $L_{41}$ in the same column as $L_{11}$, i.e., the tasks calculating the \emph{panel matrix}. In other words, there exist three data dependencies from the tasks solving $L_{21}$, $L_{31}$, and $L_{41}$ to the task factorizing $A_{11}$, denoted by three \emph{solid} directed edges from the task \textsf{Factorize(1,1)} to the tasks \textsf{Solve(2,1)}, \textsf{Solve(3,1)}, and \textsf{Solve(4,1)} individually as shown in Figure \ref{Cholesky_DAG} (see page 15). Besides the above \emph{explicit} dependencies, there exists an \emph{implicit} dependency between the task updating local matrix $A_{32}$ and the task subsequently solving $L_{32}$ on the same local matrix, denoted by the \emph{dashed} directed edge from the task \textsf{Update1(3,2)} to the task \textsf{Solve(3,2)} in Figure \ref{Cholesky_DAG}. Note that communication among tasks is not shown in Figure \ref{Cholesky_DAG}, and updating diagonal local matrices and updating non-diagonal local matrices are distinguished as \textsf{Update2()} and \textsf{Update1()} respectively due to different computation time complexity.

\section{Fundamentals: Task Dependency Set and Critical Path}

Based on the task-parallel DAG representation, next we present Task Dependency Set (TDS) and Critical Path (CP) of running parallel Cholesky, LU, and QR factorizations, where TDS contains static dependency information of parallel tasks to utilize at runtime, and CP pinpoints potential energy saving opportunities in terms of slack among the tasks.

\begin{table}
\small
\centering
\caption{Notation in Algorithms 1, 2, 3, and 4 and Henceforth.}
\label{notation1}
\begin{tabular}{|c|l|}
\hline
$task, t_1, t_2$ & One task of matrix factorizations\\
\hline
$N_{proc}$ & Square root of the total number of processes\\
\hline
$f_l$ & The lowest CPU frequency set by DVFS\\
\hline
$f_h$ & The highest CPU frequency set by DVFS\\
\hline
$f_{opt}$ & Optimal ideal frequency to finish a task with its slack eliminated\\
\hline
$ratio$ & Ratio between split durations for optimal frequency approximation\\
\hline
$\mathsf{TDS}_{in}$\textsf{($task$)} & TDS consisting of tasks that are depended by $task$ as the input\\ 
\hline
$\mathsf{TDS}_{out}$\textsf{($task$)} & TDS consisting of tasks that depend on $task$ as the input\\ 
\hline
$CritPath$ & One task trace to finish matrix factorizations with zero total slack\\
\hline
$slack$ & Time that a task can be delayed by with no overall performance loss\\
\hline
$CurFreq$ & Current CPU frequency in use\\
\hline
$DoneFlag$ & Indicator of the finish of a task\\
\hline
\end{tabular}
\normalsize
\end{table}

\subsection{Task Dependency Set}

For determining the appropriate timing for exploiting potential energy saving opportunities via DVFS, we leverage TDS analysis in our \texttt{TX} approach. Next we first formally define TDS, and then showcase how to generate two types of TDS for each task in Cholesky factorization using Algorithm 1. Producing TDS for LU and QR factorizations is similar with minor changes in Algorithm 1 per algorithmic characteristics. Table \ref{notation1} lists the notation used in the algorithms and discussion in sections 3 and 4.

\vspace{1mm}
\noindent\textsc{\textbf{Definition 2}}. Given a task $t$ of a parallel Cholesky/LU/QR factorization, data dependencies related to a data block manipulated by the task $t$ are classified as elements of two types of TDS: $\mathsf{TDS}_{in}$\textsf{($t$)} and $\mathsf{TDS}_{out}$\textsf{($t$)}, where dependencies from the data block to other tasks $t_i$ are categorized into $\mathsf{TDS}_{out}$\textsf{($t$)} and denoted as $t_i$ for short, and dependencies from other tasks $t_j$ to the data block are categorized into $\mathsf{TDS}_{in}$\textsf{($t$)} and denoted as $t_j$ for short.

\begin{figure}
\centering
\begin{tabular}{@{}p{\columnwidth}@{}}
\toprule
\textbf{Algorithm 1} ~ \textit{Task Dependency Set Generation Algorithm}
\\\midrule
\textsf{GenTDS}\textsf{(}$task$, $N_{proc}$\textsf{)}\\
\hspace{2.09mm}1: \textbf{switch} ($task$)\\
\hspace{2.09mm}2: \quad \textbf{case} \textsf{Factorize}:\\
\hspace{2.09mm}3: \quad\quad \textbf{foreach} $i < j,~1 \leq i,j \leq N_{proc}$ \textbf{do}\\
\hspace{2.09mm}4: \quad\quad\quad \textsf{insert($\mathsf{TDS}_{in}$(S($j,i$)), F($i,i$));}\\
\hspace{2.09mm}5: \quad\quad\quad \textsf{insert($\mathsf{TDS}_{out}$(F($i,i$)), S($j,i$));}\\
\hspace{2.09mm}6: \quad \textbf{case} \textsf{Update1}:\\
\hspace{2.09mm}7: \quad\quad \textbf{foreach} $i < j,~1 \leq i,j \leq N_{proc}$ \textbf{do}\\
\hspace{2.09mm}8: \quad\quad\quad \textbf{if} (\textsf{IsLastInstance($\mathsf{U_1}$($j,i$))}) \textbf{then}\\
\hspace{2.09mm}9: \quad\quad\quad\quad \textsf{insert($\mathsf{TDS}_{in}$($\mathsf{U_1}$($j,i$)), S($j,i$));}\\
10: \quad\quad\quad\quad \textsf{insert($\mathsf{TDS}_{out}$(S($j,i$)), $\mathsf{U_1}$($j,i$));}\\
11: \quad \textbf{case} \textsf{Update2}:\\
12: \quad\quad \textbf{foreach} $1 \leq i \leq N_{proc}$ \textbf{do}\\
13: \quad\quad\quad \textbf{if} (\textsf{IsLastInstance($\mathsf{U_2}$($i,i$))}) \textbf{then}\\
14: \quad\quad\quad\quad \textsf{insert($\mathsf{TDS}_{in}$(F($i,i$)), $\mathsf{U_2}$($i,i$));}\\
15: \quad\quad\quad\quad \textsf{insert($\mathsf{TDS}_{out}$($\mathsf{U_2}$($i,i$)), F($i,i$));}\\
16: \quad \textbf{case} \textsf{Solve}:\\
17: \quad\quad \textbf{foreach} $1 \leq i < j \leq N_{proc}$ \textbf{do}\\
18: \quad\quad\quad \textbf{foreach} $j < k \leq N_{proc}$ \textbf{do}\\
19: \quad\quad\quad\quad \textsf{insert($\mathsf{TDS}_{in}$($\mathsf{U_1}$($k,j$)), S($j,i$));}\\
20: \quad\quad\quad\quad \textsf{insert($\mathsf{TDS}_{out}$(S($j,i$)), $\mathsf{U_1}$($k,j$));}\\
21: \quad\quad\quad \textbf{foreach} $i < k < j \leq N_{proc}$ \textbf{do}\\
22: \quad\quad\quad\quad \textsf{insert($\mathsf{TDS}_{in}$($\mathsf{U_1}$($j,k$)), S($j,i$));}\\
23: \quad\quad\quad\quad \textsf{insert($\mathsf{TDS}_{out}$(S($j,i$)), $\mathsf{U_1}$($j,k$));}\\
24: \quad\quad\quad \textsf{insert($\mathsf{TDS}_{in}$($\mathsf{U_2}$($j,j$)), S($j,i$));}\\
25: \quad\quad\quad \textsf{insert($\mathsf{TDS}_{out}$(S($j,i$)), $\mathsf{U_2}$($j,j$));}\\
26: \textbf{end switch}\\
\midrule
\end{tabular}
\end{figure}

\vspace{1mm}
\noindent\textsc{\textbf{Example}}. Consider the Cholesky factorization in Figure \ref{Cholesky_matrix}. Two TDS of each task can be generated statically per algorithmic characteristics as shown in Algorithm 1: Since the resulting local matrices of factorization tasks (e.g., $L_{11}$) are used in column-wise \emph{panel matrix} solving (e.g., solving $L_{21}$, $L_{31}$, and $L_{41}$), data dependencies from \emph{panel matrices} to factorized \emph{diagonal matrices} are included in $\mathsf{TDS}_{in}$ of tasks solving \emph{panel matrices} (e.g., $\mathsf{TDS}_{in}$\textsf{(S(2,1))}, $\mathsf{TDS}_{in}$\textsf{(S(3,1))}, and $\mathsf{TDS}_{in}$\textsf{(S(4,1))}), and $\mathsf{TDS}_{out}$ of tasks factorizing \emph{diagonal matrices} (e.g., $\mathsf{TDS}_{out}$\textsf{(F(1,1))}). Likewise $\mathsf{TDS}_{in}$ and $\mathsf{TDS}_{out}$ of other tasks holding different dependencies can be produced following Algorithm 1.

\subsection{Critical Path}

Although load balancing techniques are leveraged for distributing workloads into a cluster of computing nodes as evenly as possible, assuming that all nodes have the same hardware configuration and thus the same computation and communication capability, slack can result from the fact that different processes can be utilized unfairly due to three primary reasons: (a) imbalanced computation delay due to data dependencies among tasks, (b) imbalanced task partitioning, and (c) imbalanced communication delay. Difference in CPU utilization results in different amount of computation slack. For instance, constrained by data dependencies, the start time of processes running on different nodes differs from each other, as shown in Figure \ref{Cholesky_DAG} (see page 15) where \textsf{P1} starts earlier than the other three processes. Moreover, since the location of local matrices in the global matrix determines what types of computation are performed on the local matrices, load imbalancing from difference in task types and task amount allocated to different processes cannot be eliminated completely by the 2-D block cyclic data distribution, as shown in Figure \ref{Cholesky_DAG} where \textsf{P2} has lighter workloads compared to the other three processes. Imbalanced communication time due to different task amount among the processes further extends the difference in slack length for different processes.

Critical Path (CP) is one particular task trace from the beginning task of one run of a task-parallel application to the ending one with the total slack of zero. Any delay on tasks on the CP increases the total execution time of the application, while dilating tasks off the CP into their slack individually without further delay does not cause performance loss as a whole. Energy savings can be achieved by appropriately reducing frequency to dilate tasks off the CP into their slack as much as possible, which is referred to as the \emph{CP} approach. Numerous existing OS level solutions effectively save energy via \emph{CP-aware} analysis \cite{ipdps05a} \cite{sc06} \cite{sc07} \cite{ics09} \cite{hpcs11} \cite{csrd12a}. Figure \ref{Cholesky_DAG} highlights one CP for the provided parallel Cholesky factorization with bold edges. We next present a feasible algorithm to generate a CP in parallel Cholesky, LU, and QR factorizations via TDS analysis.

\subsection{Critical Path Generation via TDS}

\begin{figure}
\centering
\begin{tabular}{@{}p{\columnwidth}@{}}
\toprule
\textbf{Algorithm 2} ~ \textit{Critical Path Generation Algorithm via TDS Analysis}
\\\midrule
\textsf{GenCritPath}\textsf{(}$CritPath$, $task$, $N_{proc}$\textsf{)}\\
\hspace{2.09mm}1: $CritPath$ $\leftarrow$ $\varnothing$\\
\hspace{2.09mm}2: \textbf{switch} ($task$)\\
\hspace{2.09mm}3: \quad \textbf{case} \textsf{Factorize}:\\
\hspace{2.09mm}4: \quad\quad \textsf{insert($CritPath$, F($i,i$))}\\
\hspace{2.09mm}5: \quad \textbf{case} \textsf{Update1}:\\
\hspace{2.09mm}6: \quad\quad \textsf{Do Nothing}\\
\hspace{2.09mm}7: \quad \textbf{case} \textsf{Update2}:\\
\hspace{2.09mm}8: \quad\quad \textbf{if} ($t_1 \in CritPath$ \&\& $t_1 \in \mathsf{TDS}_{out}$\textsf{($\mathsf{U_2}$($i,i$))}) \textbf{then}\\
\hspace{2.09mm}9: \quad\quad\quad \textsf{insert($CritPath$, $\mathsf{U_2}$($i,i$))}\\
10: \quad \textbf{case} \textsf{Solve}:\\
11: \quad\quad \textbf{foreach} $1 \leq i < j \leq N_{proc}$ \textbf{do}\\
12: \quad\quad\quad \textbf{if} ($t_2 \in CritPath$ \&\& $t_2 \in \mathsf{TDS}_{out}$\textsf{(S($j,i$))}) \textbf{then}\\
13: \quad\quad\quad\quad \textsf{insert($CritPath$, S($j,i$))}\\
14: \textbf{end switch}\\
\midrule
\end{tabular}
\end{figure}

We can generate a CP for parallel Cholesky, LU, and QR factorizations as the basis of the \emph{CP} approach using Algorithm 2. Consider the same parallel Cholesky factorization above. The heuristic of the CP generation algorithm is as follows: (a) Each task of factorizing is included in the CP, since the local matrices to factorize are always updated last, compared to other local matrices in the same row of the global matrix, and the outcome of factorizing is required in future computation. In other words, the task of factorizing serves as a transitive step that cannot be executed in parallel together with other tasks; (b) each task of \textsf{Update1()} is excluded from the CP, since it does not have direct dependency relationship with any tasks of factorizing, which are already included in the CP; (c) regarding \textsf{Update2()}, we select the ones that are directly depended by the tasks of factorizing on the same local matrix into the CP; (d) we choose the tasks of solving that are directly depended by \textsf{Update2()} (or directly depends on \textsf{Factorize()}, not shown in the algorithm) into the CP. Note that CP can also be identified using methods other than TDS analysis \cite{ipdps05a} \cite{hpcs11} \cite{csrd12a}.

\section{TX: Energy Efficient Race-to-halt DVFS Scheduling}

In this section, we present in detail three energy efficient DVFS scheduling approaches for parallel Cholesky, LU, and QR factorizations individually: the \emph{SC} approach, the \emph{CP} approach, and our \texttt{TX} approach. We further demonstrate that \texttt{TX} is able to save energy substantially compared to the other two solutions, since via \emph{TDS-based race-to-halt}, it circumvents the defective prediction mechanism employed by the \emph{CP} approach at OS level, and further saves energy from possible load imbalance. Moreover, we formally prove that \texttt{TX} is comparable to the \emph{CP} approach in energy saving capability under the circumstance of current CMOS technologies.

\subsection{Custom Functions}

In Algorithms 1, 2, 3, and 4, nine custom functions are introduced for readability: \textsf{insert($\mathsf{TDS}$($t_1$), $t_2$)}, \textsf{delete($\mathsf{TDS}$($t_1$), $t_2$)}, \textsf{SetFreq()}, \textsf{IsLastInstance()}, \textsf{Send()}, \textsf{Recv()}, \textsf{IsFinished()}, \textsf{GetSlack()}, and \textsf{GetOptFreq()}. The implementation of \textsf{insert()} and \textsf{delete()} is straightforward: Add task $t_2$ into the TDS of task $t_1$, and remove $t_2$ from the TDS of $t_1$. \textsf{SetFreq()} is a wrapper of handy DVFS APIs that set specific frequencies, and \textsf{Send()} and \textsf{Recv()} are wrappers of MPI communication routines that send and receive flag messages among tasks respectively. \textsf{IsLastInstance()} is employed to determine if the current task is the last instance of the same type of tasks manipulating the same data block, and \textsf{IsFinished()} is employed to determine if the current task is finished: Both are easy to implement at library level. \textsf{GetSlack()} and \textsf{GetOptFreq()} are used to get slack of a task, and calculate the optimal CPU frequency to dilate a task into its slack as much as possible, respectively. Implementing \textsf{GetSlack()} and \textsf{GetOptFreq()} can be highly non-trivial. Specifically, \textsf{GetSlack()} calculates slack of a task off the CP from the difference between the latest and the earliest end time of the task. \textsf{GetOptFreq()} calculates the optimal ideal frequency to eliminate slack of a task from the mapping between frequency and execution time for each type of tasks.

\subsection{Scheduled Communication Approach}

One effective and straightforward solution to save energy for task-parallel applications is to set CPU frequency to high during computation, while set it to low during communication, given the fact that large-message communication is not bound by CPU performance while the computation is, so the peak CPU performance is not necessary during the communication. Although substantial energy savings can be achieved from this Scheduled Communication (\emph{SC}) approach \cite{sc06} \cite{sc07}, it leaves potential energy saving opportunities from other types of slack (e.g., see slack shown in Figure \ref{Cholesky_DAG} on page 15) untapped. More energy savings can be fulfilled via fine-grained analysis of execution characteristics of HPC applications, in particular during non-communication. Next we present two well-designed approaches that take advantage of computation slack to further gain energy savings. Note since the \emph{SC} approach does not conflict with solutions exploiting slack from non-communication, it thus can be incorporated with the next two solutions seamlessly to maximize energy savings.

\subsection{Critical Path Approach vs. TX Approach}

\begin{figure}
\centering
\includegraphics[width=5.4in]{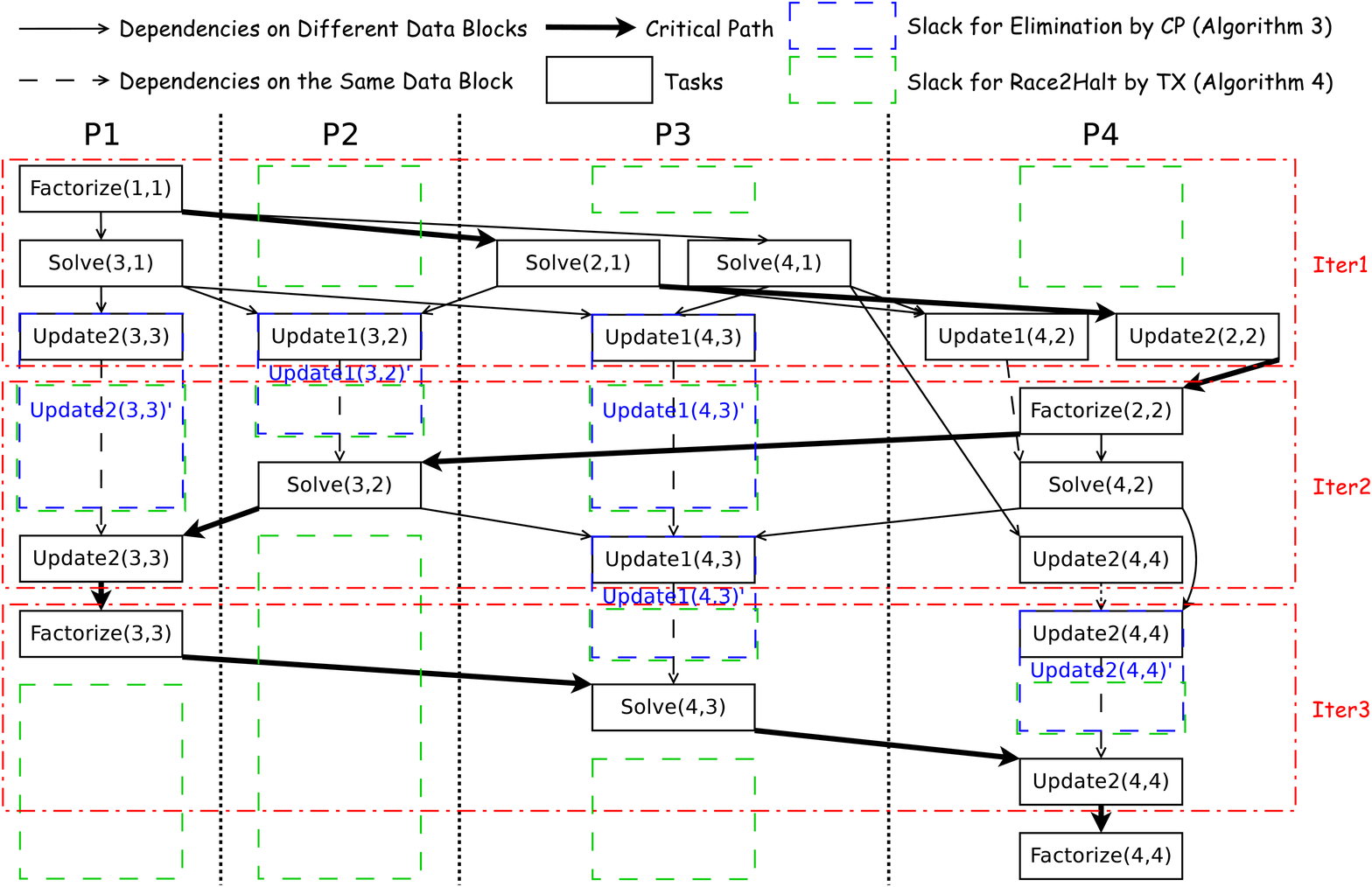}\\
\caption{DAG Representation of Task and Slack Scheduling of CP and TX Approaches for the 4 $\times$ 4 Blocked Cholesky Factorization in Figure \ref{Cholesky_matrix} on a 2 $\times$ 2 Process Grid Using 2-D Block Cyclic Data Distribution.}
\label{Cholesky_DAG}
\end{figure}

Given a detected CP (e.g., via static analysis \cite{ipdps05a} or local information analysis \cite{ics09}) for task-parallel applications, the Critical Path (\emph{CP}) approach saves energy as shown in Algorithm 3: For all tasks on the CP, the CPU operating frequency is set to the highest for attaining the peak CPU performance, while for tasks not on the CP with the total slack larger than zero (e.g., tasks with no outgoing \emph{explicit} data dependencies in Figure \ref{Cholesky_DAG}), lowering frequency appropriately is performed to dilate the tasks into their slack as much as possible, without incurring performance loss of the applications. Due to the discrete domain of available CPU frequencies defined for DVFS, if the calculated optimal frequency that can eliminate slack lies in between two available neighboring frequencies, the two frequencies can be employed to approximate it by calculating a ratio of durations operating at the two frequencies. The two frequencies are then assigned to the durations separately based on the ratio. Lines 7-9 in Algorithm 3 sketch the frequency approximation method \cite{tpds04} \cite{ics09}. The ratio of split frequencies is calculated via prior knowledge of the mapping between frequency and execution time of different types of tasks. Note that we denote the two neighboring available frequencies of $f_{opt}$ as $\lfloor{f_{opt}}\rfloor$ and $\lceil{f_{opt}}\rceil$.

\begin{figure}
\centering
\begin{tabular}{@{}p{\columnwidth}@{}}
\toprule
\textbf{Algorithm 3} ~ \textit{DVFS Scheduling Algorithm Using CP}
\\\midrule
\textsf{DVFS\_CP}\textsf{(}$CritPath$, $task$, $FreqSet$\textsf{)}\\
\hspace{2.09mm}1: \textbf{if} ($task \in CritPath$ $||$ $\mathsf{TDS}_{out}$\textsf{($task$)} != $\varnothing$) \textbf{then}\\
\hspace{2.09mm}2: \quad \textsf{SetFreq($f_h$)}\\
\hspace{2.09mm}3: \textbf{else}\\
\hspace{2.09mm}4: \quad $slack$ $\leftarrow$ \textsf{GetSlack($task$)}\\
\hspace{2.09mm}5: \quad \textbf{if} ($slack > 0$) \textbf{then}\\
\hspace{2.09mm}6: \quad\quad $f_{opt}$ $\leftarrow$ \textsf{GetOptFreq($task$, $slack$)}\\
\hspace{2.09mm}7: \quad \textbf{if} ($f_l \leq f_{opt} \leq f_h$) \textbf{then}\\
\hspace{2.09mm}8: \quad\quad \textbf{if} ($f_{opt} \notin FreqSet$) \textbf{then}\\
\hspace{2.09mm}9: \quad\quad\quad \textsf{SetFreq($\lfloor{f_{opt}}\rfloor$, $\lceil{f_{opt}}\rceil$, $ratio$)}\\
10: \quad\quad \textbf{else} \textsf{SetFreq($f_{opt}$)}\\
11: \quad \textbf{else if} ($f_{opt} < f_l$) \textbf{then}\\
12: \quad\quad \textsf{SetFreq($f_l$)}\\
13: \textbf{end if}\\
\midrule
\end{tabular}
\end{figure}
\begin{figure}
\centering
\begin{tabular}{@{}p{\columnwidth}@{}}
\toprule
\textbf{Algorithm 4} ~ \textit{DVFS Scheduling Algorithm Using TX}
\\\midrule
\textsf{DVFS\_TX}\textsf{(}$task$, $CurFreq$\textsf{)}\\
\hspace{2.09mm}1: \textbf{while} ($\mathsf{TDS}_{in}$\textsf{($task$)} != $\varnothing$) \textbf{do}\\
\hspace{2.09mm}2: \quad \textbf{if} ($CurFreq$ != $f_l$) \textbf{then}\\
\hspace{2.09mm}3: \quad\quad \textsf{SetFreq($f_l$)}\\
\hspace{2.09mm}4: \quad \textbf{if} (\textsf{Recv($DoneFlag$, $t_1$)}) \textbf{then}\\
\hspace{2.09mm}5: \quad\quad \textsf{delete($\mathsf{TDS}_{in}$($task$), $t_1$)}\\
\hspace{2.09mm}6: \textbf{end while}\\
\hspace{2.09mm}7: \textsf{SetFreq($f_h$)}\\
\hspace{2.09mm}8: \textbf{if} (\textsf{IsFinished($task$)}) \textbf{then}\\
\hspace{2.09mm}9: \quad \textbf{foreach} $t_2 \in \mathsf{TDS}_{out}$\textsf{($task$)} \textbf{do}\\
10: \quad\quad \textsf{Send($DoneFlag$, $t_2$)}\\
11: \quad \textsf{SetFreq($f_l$)}\\
12: \textbf{end if}\\
\midrule
\end{tabular}
\end{figure}

Different from the \emph{CP} approach that reduces CPU frequency for tasks off the CP to eliminate slack for saving energy without performance loss, \texttt{TX} employs the \emph{race-to-halt} mechanism that leverages two TDS of each task ($\mathsf{TDS}_{in}$ and $\mathsf{TDS}_{out}$) to determine the timing of \emph{race} and \emph{halt}, as shown in Algorithm 4. Respecting data dependencies, one \emph{dependent} task cannot start until the finish of its \emph{depended} task. \texttt{TX} keeps the \emph{dependent} task staying at the lowest frequency, i.e., \emph{halt}, until all its \emph{depended} tasks have finished when it may start, and then allows the \emph{dependent} task to work at the highest frequency to complete as soon as possible, i.e., \emph{race}, before being switched back to the low-power state. Upon completion, a task sends a $DoneFlag$ to all its \emph{dependent} tasks to notify them that data needed has been processed and ready for use. A \emph{dependent} task is retained at the lowest frequency while waiting for $DoneFlags$ from all its \emph{depended} tasks, and removes the dependency to a \emph{depended} task from its $\mathsf{TDS}_{in}$, once a $DoneFlag$ from the \emph{depended} task is received.

\vspace{1mm}
\noindent\textsc{\textbf{Defective Prediction Mechanism}}. Although effective, the OS level \emph{CP} approach essentially depends on the workload prediction mechanism: Execution characteristics in the upcoming interval can be predicted using prior execution information, e.g., execution traces in recent intervals. However, this prediction mechanism may not necessarily be reliable and lightweight: (a) For applications with variable execution characteristics, such as dense matrix factorizations. Execution time of iterations of the core loop shrinks as the remaining unfinished matrices become smaller. Consequently dynamic prediction on execution characteristics such as task runtime and workload distribution can be inaccurate, which thus leads to error-prone slack estimation; (b) for applications with random execution patterns, such as applications relying on random numbers that could lead to variation of control flow at runtime, which can be difficult to capture. Since the predictor needs to determine reproducible execution patterns at the beginning of one execution, it can be costly for obtaining accurate prediction results in both cases above. Given the fact that no energy savings can be fulfilled until the prediction phase is finished, considerable potential energy savings may be wasted for accurate but lengthy prediction as such at OS level.

There exist numerous studies on history-based workload prediction, which can be generally considered as two variants: the simplest and mostly commonly-used prediction algorithm PAST \cite{osdi94} and its enhanced algorithms \cite{cases03} \cite{date04} \cite{sc05c} \cite{icpp07}, where PAST works as follows:

\vspace{-5mm}
\begin{equation}
\label{past_algorithm}
W_{i+1}' = W_i
\vspace{-1mm}
\end{equation}

\noindent where $W_{i+1}'$ is the next executed workload to predict, and $W_i$ is the current measured workload. For applications with stable or slowly-varying execution patterns, the straightforward PAST algorithm can work well with little performance loss and high accuracy. It is however not appropriate for handling a considerable amount of variation in execution patterns. Many enhanced algorithms have been proposed to produce more accurate workload prediction for such applications. For instance, the RELAX algorithm employed in CPU MISER \cite{icpp07} exploits both prior predicted profiles and current runtime measured profiles as follows:

\vspace{-5mm}
\begin{equation}
\label{relax_algorithm}
W_{i+1}' = (1 - \lambda) W_i' + \lambda W_i
\vspace{-1mm}
\end{equation}

\noindent where $\lambda$ is a relaxation factor for adjusting the extent of dependent information of the current measurement. This enhanced prediction mechanism can also be error-prone and costly for parallel Cholesky, LU, and QR factorizations due to the use of 2-D block cyclic data distribution. As shown in Figure \ref{Cholesky_DAG}, across loop iterations (highlighted by red dashed boxes) different types of tasks can be distributed to a process. For instance, in the first iteration, $\mathsf{P_1}$ is assigned three tasks \textsf{Factorize(1,1)}, \textsf{Solve(3,1)}, and \textsf{Update2(3,3)}, while in the second iteration, it is only assigned one task \textsf{Update2(3,3)}. Although empirically for a large global matrix, tasks are distributed to one specific process alternatingly (e.g., \textsf{Factorize(i,i)} ($\mathsf{i = 1, 3, 5, \dots, 2s+1, s \geq 0}$) can be all distributed to \textsf{P1}), the RELAX algorithm needs to adjust the value of $\lambda$ alternatingly as well, which can be very costly and thus diminish potential energy savings. Length variation of iterations due to the shrinking remaining unfinished matrices further brings complexity to workload prediction.

\texttt{TX} successfully circumvents the shortcomings from the OS level workload prediction, by leveraging the TDS-based \emph{race-to-halt} strategy. \texttt{TX} essentially differs from the \emph{CP} approach since it allows tasks to complete as soon as possible before entering the low-power state, and thus does not require any workload prediction mechanism. \texttt{TX} works at library level and thus benefits from: (a) obtaining application-specific knowledge easily that can be utilized to determine accurate timing for DVFS, and (b) restricting the scope of source modification and recompilation to library level, compared to application level solutions.

\vspace{1mm}
\noindent\textsc{\textbf{Potential Energy Savings from Load Imbalance}}. Due to the existence of load imbalance regardless of load balancing techniques applied, and also data dependencies among inter-process parallel tasks, not all tasks can start to work and finish at the same time, as shown in Figure \ref{Cholesky_DAG}. At OS level, the \emph{SC} approach and the \emph{CP} approach only work when tasks are being executed, which leaves potential energy savings untapped during the time otherwise. Specifically, the \emph{SC} approach switches frequency to high and low at the time when computation and communication tasks start respectively, and do nothing during the time other than computation and communication. Likewise, the \emph{CP} approach assigns appropriate frequencies for tasks on/off the CP individually. Therefore due to its nature of keeping the peak CPU performance for the tasks on the CP and dilating the tasks off the CP into its slack via frequency reduction, switching frequency is not feasible when for one process no tasks have started or all tasks have already finished.

Due to its nature of \emph{race-to-halt}, \texttt{TX} can start to save energy even before any tasks in a process are executed, and after all tasks in a process have finished while there exist unfinished tasks in other processes. In Figure \ref{Cholesky_DAG}, the durations only covered by green dashed boxes highlights the additional energy savings fulfilled by \texttt{TX}, where due to data dependencies, a task cannot start yet while waiting for its depended task to finish first, or all tasks in one process have already finished while some tasks in other processes are still running.

\begin{table}
\small
\centering
\caption{Notation in Energy Saving Analysis.}
\label{notation2}
\begin{tabular}{|c|l|}
\hline
$E$ & The total nodal energy consumption of all components\\
\hline
$P$ & The total nodal power consumption of all components\\
\hline
$P_{dynamic}$ & Dynamic power consumption in the running state\\
\hline
$P_{leakage}$ & Static/leakage power consumption in any states\\
\hline
$T$ & Execution time of a task at the peak CPU performance\\
\hline
$T'$ & Slack of executing a task at the peak CPU performance\\
\hline
$A$ & Percentage of active gates in a CMOS-based chip\\
\hline
$C$ & The total capacitive load in a CMOS-based chip\\
\hline
$f$ & Current CPU operating frequency\\
\hline
$V$ & Current CPU supply voltage\\
\hline
$V'$ & Supply voltage of components other than CPU\\
\hline
$I_{sub}$ & CPU subthreshold leakage current\\
\hline
$I_{sub}'$ & non-CPU component subthreshold leakage current\\
\hline
$f_m$ & Available optimal frequency assumed to eliminate $T'$\\
\hline
$V_h$ & The highest supply voltage corresponding to $f_h$ set by DVFS\\
\hline
$V_l$ & The lowest supply voltage corresponding to $f_l$ set by DVFS\\
\hline
$V_m$ & Supply voltage corresponding to $f_m$ set by DVFS\\
\hline
$n$ & Ratio between original runtime and slack of a task\\
\hline
\end{tabular}
\normalsize
\end{table}


\vspace{1mm}
\noindent\textsc{\textbf{Energy Saving Capability Analysis}}. Next we formally prove that compared to the classic \emph{CP} approach, \texttt{TX} is comparable to it in energy saving capability. Given the following two energy saving strategies, towards a task $t$ with an execution time $T$ and slack $T'$ at the peak CPU performance, we calculate the total nodal system energy consumption for both strategies, i.e., $E(\mathsf{S_1})$ and $E(\mathsf{S_2})$ formally below:

\begin{itemize}
\item \textsf{Strategy I (Race-to-halt)}: Execute $t$ at the highest frequency $f_h$ until the finish of $t$ and then switch to the lowest frequency $f_l$, i.e., run in $T$ at $f_h$ and then run in $T'$ at $f_l$;
\item \textsf{Strategy II (CP-aware)}: Execute $t$ at the optimal frequency $f_m$ with which $T'$ is eliminated, i.e., run in $T + T'$ at $f_m$ (without loss of generality, assume $T'$ can be eliminated using an available frequency $f_m$ without frequency approximation).
\end{itemize}

For simplicity, let us assume the tasks for the use of DVFS are compute-intensive (memory-intensive tasks can be discussed with minor changes in the model), i.e., $T + T' = nT$, when $f_m = \frac{1}{n}f_h$, where $1 \leq n \leq \frac{f_h}{f_l}$. Considering the nodal power consumption $P$, we model it formally as follows:

\vspace{-3mm}
\begin{equation}
\label{node_power1}
P = P_{dynamic}^{CPU} + P_{leakage}^{CPU} + P_{leakage}^{other}
\end{equation}

\vspace{-5.5mm}
\begin{equation}
\label{dynamic_power}
P_{dynamic} = ACfV^2
\end{equation}

\vspace{-5.5mm}
\begin{equation}
\label{leakage_power}
P_{leakage} = I_{sub}V
\vspace{0mm}
\end{equation}

Then, substituting Equations \ref{dynamic_power} and \ref{leakage_power} into Equation \ref{node_power1} yields:

\vspace{-3mm}
\begin{equation}
\label{node_power2}
P = ACfV^2 + I_{sub}V + I_{sub}'V'
\vspace{0mm}
\end{equation}

In our scenario, $P_{leakage}^{other} = I_{sub}'V'$ is independent of CPU frequency and voltage scaling, and thus can be regarded as a constant in Equation \ref{node_power2}, so we denote $P_{leakage}^{other}$ as $P_c$ for short. Further, although subthreshold leakage current $I_{sub}$ has an exponential relationship with threshold voltage, results presented in \cite{edl04} indicate that $I_{sub}$ converges to a constant after a certain threshold voltage value. Without loss of generality, we treat $P_{leakage}^{CPU} = I_{sub}V$ as a function of supply voltage $V$ only. Thus, we model the nodal energy consumption $E_{node}$ for both strategies individually below:

\vspace{-3mm}
\begin{align}
\label{node_energy_strategy1}
&E(\mathsf{S_1}) = \overline{P(\mathsf{S_1})} \times T + \overline{P'(\mathsf{S_1})} \times T'\nonumber\\
&= (ACf_hV_h^2 + I_{sub}V_h + P_c) T + (ACf_lV_l^2 + I_{sub}V_l + P_c) T'\nonumber\\
&= AC (f_hV_h^2T + f_lV_l^2T') + I_{sub}(V_hT + V_lT') + P_c (T + T')
\end{align}

\vspace{-5mm}
\begin{align}
\label{node_energy_strategy2}
&E(\mathsf{S_2}) = \overline{P(\mathsf{S_2})} \times (T + T')\nonumber\\
&= (ACf_mV_m^2 + I_{sub}V_m + P_c) (T + T')\nonumber\\
&= ACf_mV_m^2 (T + T') + I_{sub}V_m (T + T') + P_c (T + T')
\end{align}

We obtain the difference between energy costs of both strategies by dividing Equation \ref{node_energy_strategy2} by Equation \ref{node_energy_strategy1}:

\vspace{-4mm}
\begin{equation}
\label{node_energy_savings1}
\frac{E(\mathsf{S_2})}{E(\mathsf{S_1})} = \frac{ACf_mV_m^2 (T + T') + I_{sub}V_m (T + T') + P_c (T + T')}{AC (f_hV_h^2T + f_lV_l^2T') + I_{sub}(V_hT + V_lT') + P_c (T + T')}
\end{equation}

Substituting the assumption that $T' = (n-1) T$ and $f_m = \frac{1}{n}f_h$ into both numerator and denominator yields the following simplified formula:

\vspace{-4mm}
\begin{align}
\label{node_energy_savings2}
\frac{E(\mathsf{S_2})}{E(\mathsf{S_1})} &= \frac{AC\frac{1}{n}f_hV_m^2 \left(1 + \left(n-1\right)\right) + I_{sub}V_m \left(1 + \left(n-1\right)\right) + P_c \left(1 + \left(n-1\right)\right)}{AC \left(f_hV_h^2 + f_lV_l^2\left(n-1\right)\right) + I_{sub}\left(V_h + V_l\left(n-1\right)\right) + P_c \left(1 + \left(n-1\right)\right)}\nonumber\\
&= \frac{ACf_hV_m^2 + nI_{sub}V_m + nP_c}{AC \left(f_hV_h^2 + f_lV_l^2\left(n-1\right)\right) + I_{sub}\left(V_h + V_l\left(n-1\right)\right) + nP_c}
\end{align}

In Equation \ref{node_energy_savings2}, the denominator is a function of the variable $n$ only. It is clear that it is a monotonically increasing function for $n$, whose minimum value is attained when $n = 1$, i.e., when slack $T'$ equals $0$. Given the fact that supply voltage has a positive correlation with (not strictly proportional to) operating frequency, scaling up/down frequency results in voltage up/down accordingly as shown in Table \ref{frequency_voltage_pairs}. Therefore for the numerator of Equation \ref{node_energy_savings2}, the greater $n$ is, the smaller $f_m$ and $V_m$ are, provided $f_m = \frac{1}{n}f_h$ and the above fact. It is thus complicated to determine the monotonicity of the numerator. As a matter of fact, state-of-the-art CMOS technologies allow insignificant variation of voltage as frequency scales (see Table \ref{frequency_voltage_pairs}). Consequently the term $ACf_hV_m^2$ within the numerator does not decrease much together with the increase of $n$. Moreover, the ratio between the highest and the lowest frequencies determines the upper bound of $n$ ($1 \leq n \leq \frac{f_h}{f_l}$), so the other two terms within the numerator cannot increase significantly as $n$ goes up.

\vspace{1mm}
\noindent\textsc{\textbf{Example}}. From the operating points of various processors shown in Table \ref{frequency_voltage_pairs}, we can calculate numerical energy savings for different values of $n$ and a specific processor configuration that quantify the energy efficiency of both energy saving strategies. For the AMD Opteron 2218 processor, given a task with the execution time $T$ and slack $0.25T$, i.e., $n = 1.25$, for eliminating the slack, 1.8 GHz is adopted as the operating frequency for running the task, and thus the numerator of Equation \ref{node_energy_savings2} equals $AC\left(2.4\times1.25^2+\left(1.25-1\right)\times1.0\times1.1^2\right) + I_{sub}\left(1.25+\left(1.25-1\right)\times1.1\right) + 1.25P_c = 4.0525AC + 1.525I_{sub} + 1.25P_c$, while the denominator of Equation \ref{node_energy_savings2} equals $AC\times2.4\times1.15^2 + 1.25\times1.15I_{sub} + 1.25P_c = 3.174AC + 1.4375I_{sub} + 1.25P_c$. We can see that the coefficients of the corresponding term $P_c$ between the numerator and the denominator are the same, and the coefficients of both corresponding terms $AC$ and $I_{sub}$ between the numerator and the denominator do not differ much. Therefore the result of $\frac{E(\mathsf{S_2})}{E(\mathsf{S_1})}$ would be a value close to 1, which means \textsf{Strategy I} is comparable to \textsf{Strategy II} in energy efficiency, regardless of the defective prediction mechanism.

\begin{table}
\small
\centering
\caption{Frequency-Voltage Pairs for Different Processors (Unit: Frequency (GHz), Voltage (V)).}
\label{frequency_voltage_pairs}
\begin{tabular}{|c|c|c|c|c|c|c|c|c|c|c|}
\hline
\multirow{3}{*}{\begin{turn}{-90}~Gear\end{turn}} & \multicolumn{2}{|c|}{\multirow{2}{*}{AMD}} & \multicolumn{2}{|c|}{AMD Opteron} & \multicolumn{2}{|c|}{\multirow{2}{*}{AMD}} & \multicolumn{2}{|c|}{\multirow{2}{*}{Intel}} & \multicolumn{2}{|c|}{\multirow{2}{*}{Intel Core}}\\
& \multicolumn{2}{|c|}{\multirow{2}{*}{Opteron 2380}} & \multicolumn{2}{|c|}{846 and AMD} & \multicolumn{2}{|c|}{\multirow{2}{*}{Opteron 2218}} & \multicolumn{2}{|c|}{\multirow{2}{*}{Pentium M}} & \multicolumn{2}{|c|}{\multirow{2}{*}{i7-2760QM}}\\
& \multicolumn{2}{|c|}{} & \multicolumn{2}{|c|}{Athlon64 3200+} & \multicolumn{2}{|c|}{} & \multicolumn{2}{|c|}{} & \multicolumn{2}{|c|}{}\\
\cline{2-11}
& Freq. & Volt. & Freq. & Volt. & Freq. & Volt. & Freq. & Volt. & Freq. & Volt.\\
\hline
0 & 2.5 & 1.300 & 2.0 & 1.500 & 2.4 & 1.250 & 1.4 & 1.484 & 2.4 & 1.060\\
\hline
1 & 1.8 & 1.200 & 1.8 & 1.400 & 2.2 & 1.200 & 1.2 & 1.436 & 2.0 & 0.970\\
\hline
2 & 1.3 & 1.100 & 1.6 & 1.300 & 1.8 & 1.150 & 1.0 & 1.308 & 1.6 & 0.890\\
\hline
3 & 0.8 & 1.025 & 0.8 & 0.900 & 1.0 & 1.100 & 0.8 & 1.180 & 0.8 & 0.760\\
\hline
\end{tabular}
\normalsize
\end{table}

\section{Implementation and Evaluation}

We have implemented \texttt{TX}, and for comparison purposes, the library level \emph{SC} approach to evaluate the effectiveness of \texttt{TX} to save energy during non-communication slack. For comparing with the OS level \emph{SC} and \emph{CP} approaches, we communicated with the authors of Adagio \cite{ics09} and Fermata \cite{sc07} and received the latest version of the implementation of both. We also compare with another OS level solution CPUSpeed \cite{cpuspeed}, an interval-based DVFS scheduler that scales CPU performance according to runtime CPU utilization during the past interval. Regarding future workload prediction, essentially Adagio and Fermata leverage the PAST algorithm \cite{osdi94}, and CPUSpeed uses a prediction algorithm similar to the RELAX algorithm employed in CPU MISER \cite{icpp07}. With application-specific knowledge known, all library level solutions do not need the workload prediction mechanism. In the later discussion, we denote the above approaches as follows:

\begin{itemize}
\item \texttt{Orig}: The original runs of different-scale parallel Cholesky, LU, and QR factorizations without any energy saving approaches;
\item \texttt{SC\_lib}: The library level implementation of the \emph{SC} approach;
\item \texttt{Fermata}: The OS level implementation of the \emph{SC} approach based on the PAST workload prediction algorithm;
\item \texttt{Adagio}: The OS level implementation of the \emph{CP} approach based on the PAST workload prediction algorithm, where \texttt{Fermata} is incorporated;
\item \texttt{CPUSpeed}: The OS level implementation of the \emph{SC} approach based on a workload prediction algorithm similar to the RELAX algorithm;
\item \texttt{TX}: The library level implementation of the \emph{race-to-halt} approach based on TDS analysis, where \texttt{SC\_lib} is incorporated.
\end{itemize}

The goals of the evaluation are to demonstrate that: (a) \texttt{TX} is able to save energy effectively and efficiently for applications with variable execution characteristics such as parallel Cholesky, LU, and QR factorizations, while OS level prediction-based solutions cannot maximize energy savings, and (b) \texttt{TX} only incurs negligible performance loss, similar as the compared OS level solutions. We did not compare with application level solutions, since they essentially fulfill the same energy efficiency as library level solutions, with source modification and recompilation at application level. Working at library level, \texttt{TX} was deployed in a distributed manner to each core/process. As the additional functionality of saving energy, the implementation of \texttt{TX} was embedded into a rewritten version of ScaLAPACK \cite{scalapack}, a widely used high performance and scalable numerical linear algebra library for distributed-memory architectures. In particular, library level source modification and recompilation were conducted to the \textsf{pdpotrf()}, \textsf{pdgetrf()}, and \textsf{pdgeqrf()} routines, which perform parallel Cholesky, LU, and QR factorizations, respectively.

\begin{table}[h]
\scriptsize
\centering
\caption{Hardware Configuration for All Experiments.}
\label{hardware_configuration}
\begin{tabular}{|c|c|c|}
\hline
Cluster & HPCL & ARC\\
\hline
System Size & \multirow{2}{*}{8} & \multirow{2}{*}{108}\\
(\# of Nodes) & &\\
\hline
\multirow{2}{*}{Processor} & 2$\times$Quad-core & 2$\times$8-core\\
& AMD Opteron 2380 & AMD Opteron 6128\\
\hline
CPU Freq. & 0.8, 1.3, 1.8, 2.5 GHz & 0.8, 1.0, 1.2, 1.5, 2.0 GHz\\
\hline
Memory & 8 GB RAM & 32 GB RAM\\
\hline
\multirow{2}{*}{Cache} & 128 KB L1, 512 KB L2, & 128 KB L1, 512 KB L2,\\
& 6 MB L3 & 12 MB L3\\
\hline
Network & 1 GB/s Ethernet & 40 GB/s InfiniBand\\
\hline
\multirow{2}{*}{OS} & CentOS 6.2, 64-bit & CentOS 5.7, 64-bit\\
& Linux kernel 2.6.32 & Linux kernel 2.6.32\\
\hline
Power Meter & PowerPack & Watts up? PRO\\
\hline
\end{tabular}
\normalsize
\end{table}

\subsection{Experimental Setup}

We applied all five energy efficient approaches to the parallel Cholesky, LU, and QR factorizations with five different global matrix sizes each to assess our goals. Experiments were performed on two power-aware clusters: HPCL and ARC. Table \ref{hardware_configuration} lists the hardware configuration of the two clusters. Note that we measured the total dynamic and leakage energy consumption of distributed runs using PowerPack \cite{tpds10}, a comprehensive software and hardware framework for energy profiling and analysis of high performance systems and applications. The total of static and dynamic power consumption was measured using Watts up? PRO \cite{wattsup}. Both energy and power consumption are the total energy and power costs respectively on all involved components of one compute node, such as CPU, memory, disk, motherboard, etc. Since each set of three nodes of the ARC cluster share one power meter, power consumption measured is for the total power consumption of three nodes, while energy consumption measured is for all energy costs collected from all eight nodes of the HPCL cluster. CPU DVFS was implemented via the CPUFreq infrastructure \cite{cpufreq} that directly modifies CPU frequency system configuration files.

\subsection{Results}

In this section, we present experimental results on power, energy, and performance efficiency and trade-offs individually by comparing \texttt{TX} with the other energy saving approaches.

\begin{figure}
\centering
\includegraphics[width=3.44in]{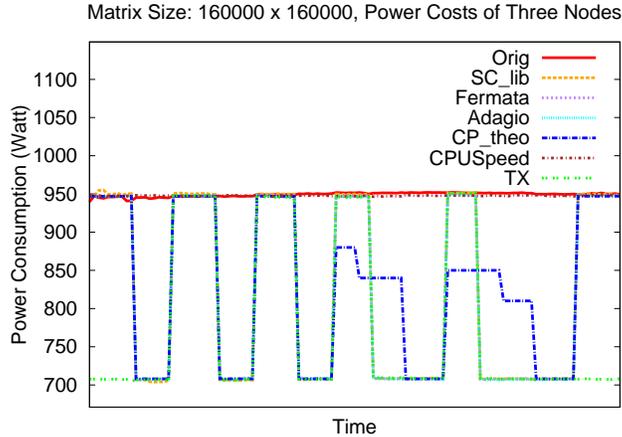}
\caption{Power Consumption of Parallel Cholesky Factorization with Different Energy Saving Approaches on the ARC Cluster using 16 $\times$ 16 Process Grid.}
\label{power_efficiency_arc}
\end{figure}

\vspace{1mm}
\noindent\textsc{\textbf{Power Savings}}. First we evaluate the capability of power savings from the five energy saving approaches for parallel Cholesky, LU, and QR factorizations on the ARC cluster (due to the similarity of results, data for parallel LU and QR factorizations is not shown), where the power consumption was measured by sampling at a constant rate through the execution of the applications. Figure \ref{power_efficiency_arc} depicts the total power consumption of three nodes (out of sixteen nodes in use) running parallel Cholesky factorization with different approaches using a 160000 $\times$ 160000 global matrix, where we select time durations of the first few iterations. Note that parallel Cholesky factorization (the core loop) performs alternating computation and communication with decreasing execution time of each iteration, as the remaining unfinished matrix shrinks. Thus we can see that for all approaches, from left to right, the durations of computation (the peak power values) decrease as the factorization proceeds.

Among the seven runs (including the theoretical one \texttt{CP\_theo}), there exist four power variation patterns: (a) \texttt{Orig} and \texttt{CPUSpeed} -- employed the same highest frequency for both computation and communication, resulting almost constant power consumption around 950 Watts; (b) \texttt{SC\_lib}, \texttt{Fermata}, and \texttt{Adagio} -- lowered down frequency during the communication, i.e., the five low-power durations around 700 Watts, and resumed the peak CPU performance during the computation; (c) \texttt{CP\_theo} -- not only scheduled low power states for communication, but also slowed down computation to eliminate computation slack -- this is a theoretical value curve instead of real measurement, which is how OS level approaches such as \texttt{Adagio} is supposed to save more power as a \emph{CP-aware} approach based on accurate workload prediction; and (d) \texttt{TX} -- employed the \emph{race-to-halt} strategy to lower down CPU performance for all durations other than computation.

Specifically, upon the prediction algorithm that inspects dynamic prior CPU utilization periodically for workload prediction, \texttt{CPUSpeed} failed to produce accurate prediction and scale CPU power states accordingly: It kept the peak CPU performance all the time. Either relying on knowing application characteristics (\texttt{SC\_lib}) or detecting MPI communication calls (\texttt{Fermata} and \texttt{Adagio}), all three approaches can identify durations of communication and apply DVFS decisions accordingly. As discussed earlier, solutions that only slow down CPU during communication are semi-optimal. \texttt{Adagio} and \texttt{TX} are expected to utilize computation slack for achieving additional energy savings. Due to the defective OS level prediction mechanism, \texttt{Adagio} failed to predict behavior of future tasks and calculate computation slack accurately. Consequently no low-power states were switched to during computation for \texttt{Adagio}. In contrast, we provide a theoretical value curve \texttt{CP\_theo} that calculates computation slack effectively, and lower power states were switched to eliminate the slack, i.e., the four medium-power durations around 850 Watts during the third and the fourth computation as the blue line shows. Different from the solutions that save energy via slack reclamation, \texttt{TX} relies on the \emph{race-to-halt} mechanism where computation is conducted at the peak CPU performance and the lowest CPU frequency is employed immediately after the computation. Therefore during computation slack, we can observe low-power states were switched to by \texttt{TX}. Moreover, the nature of \emph{race-to-halt} also guarantees no high-power states are employed during the waiting durations resulting from data dependencies and load imbalance, i.e., the two low-power durations in green where the application starts and ends. This indicates that \texttt{TX} is able to gain additional energy savings that all other approaches cannot exploit: Processes have to stay at the high-power state at the beginning/ending of the execution.

\begin{figure*}
\vspace{-12mm}
\centering
\begin{tabular}{@{}ccc@{}}
\hspace{-23mm}\includegraphics[width=.45\textwidth]{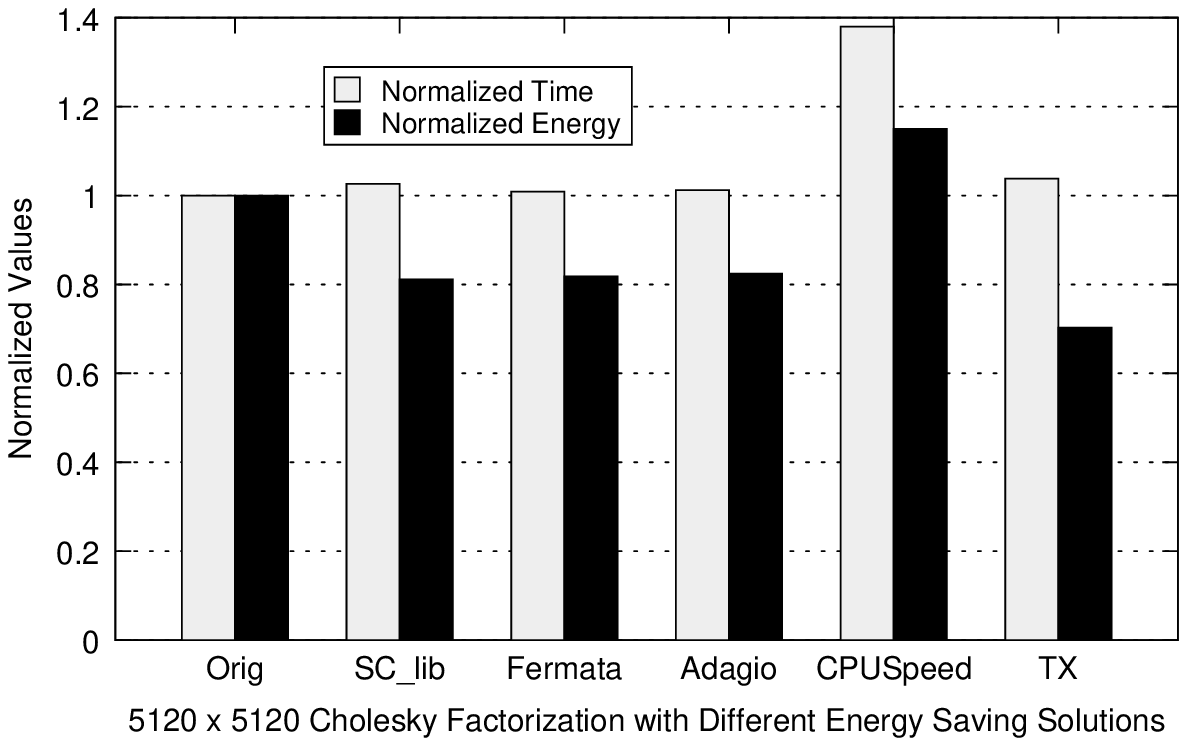} &
\hspace{-7mm}\includegraphics[width=.45\textwidth]{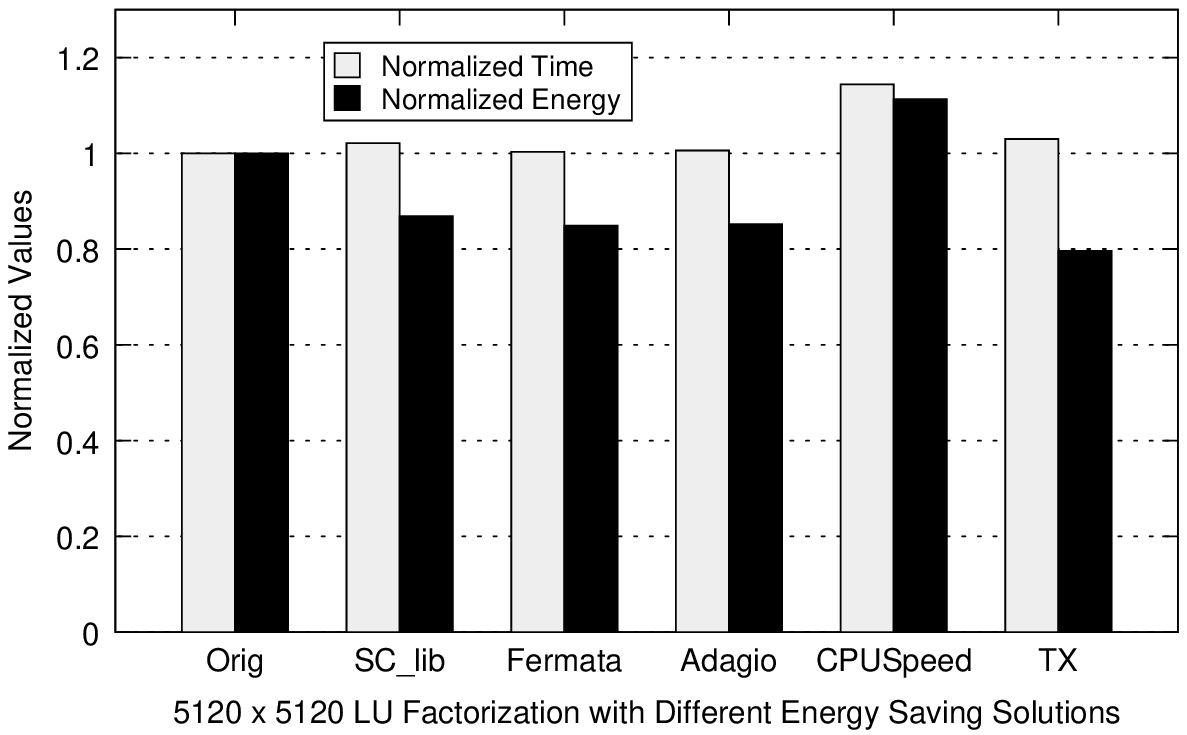} &
\hspace{-7mm}\includegraphics[width=.45\textwidth]{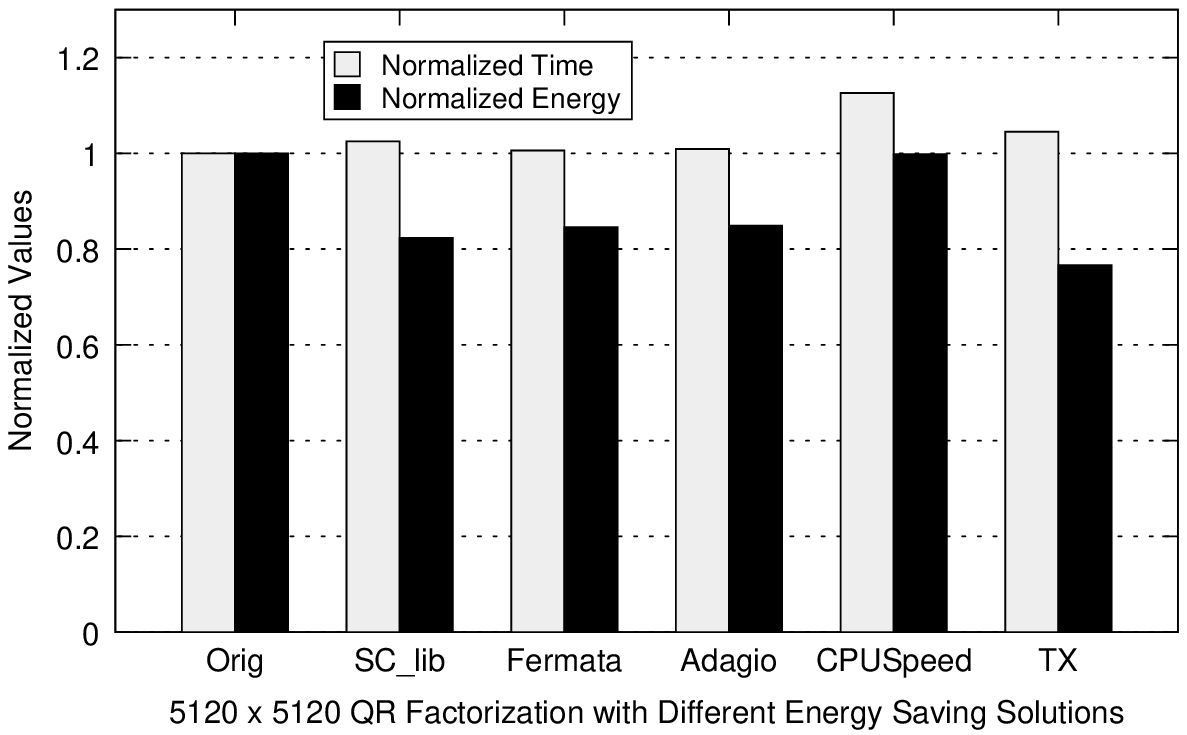}\\
\hspace{-22mm}\includegraphics[width=.46\textwidth]{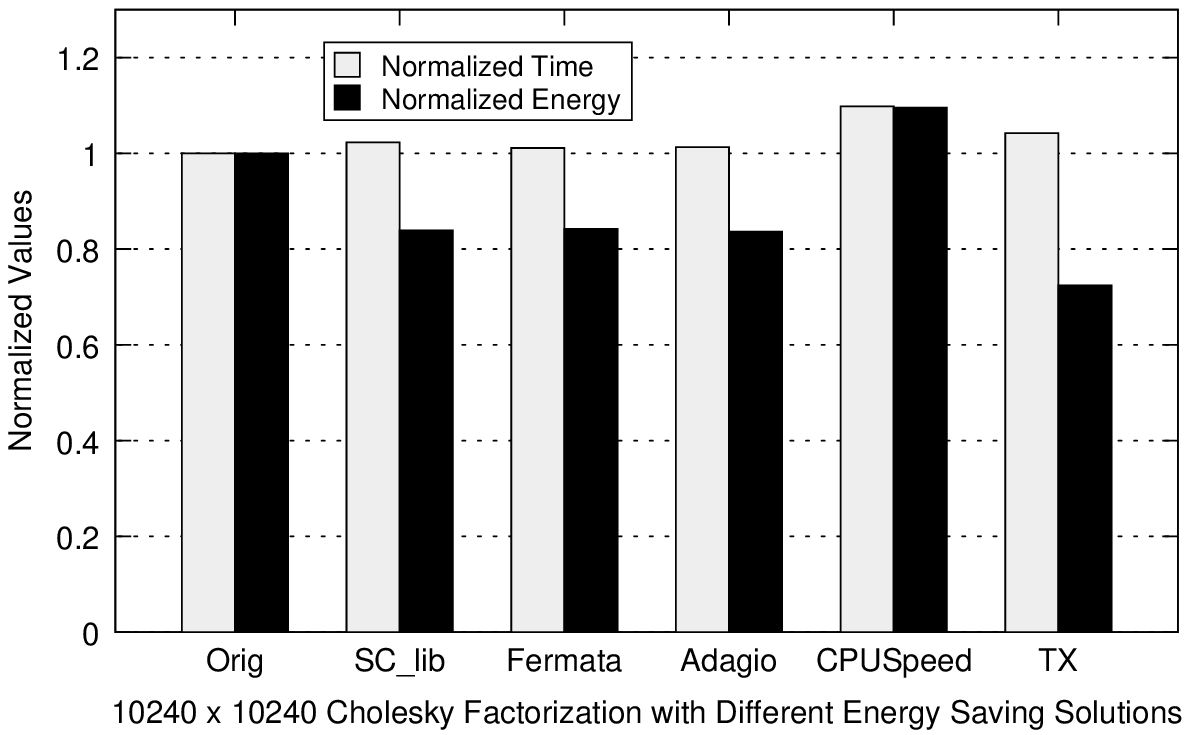} &
\hspace{-7mm}\includegraphics[width=.46\textwidth]{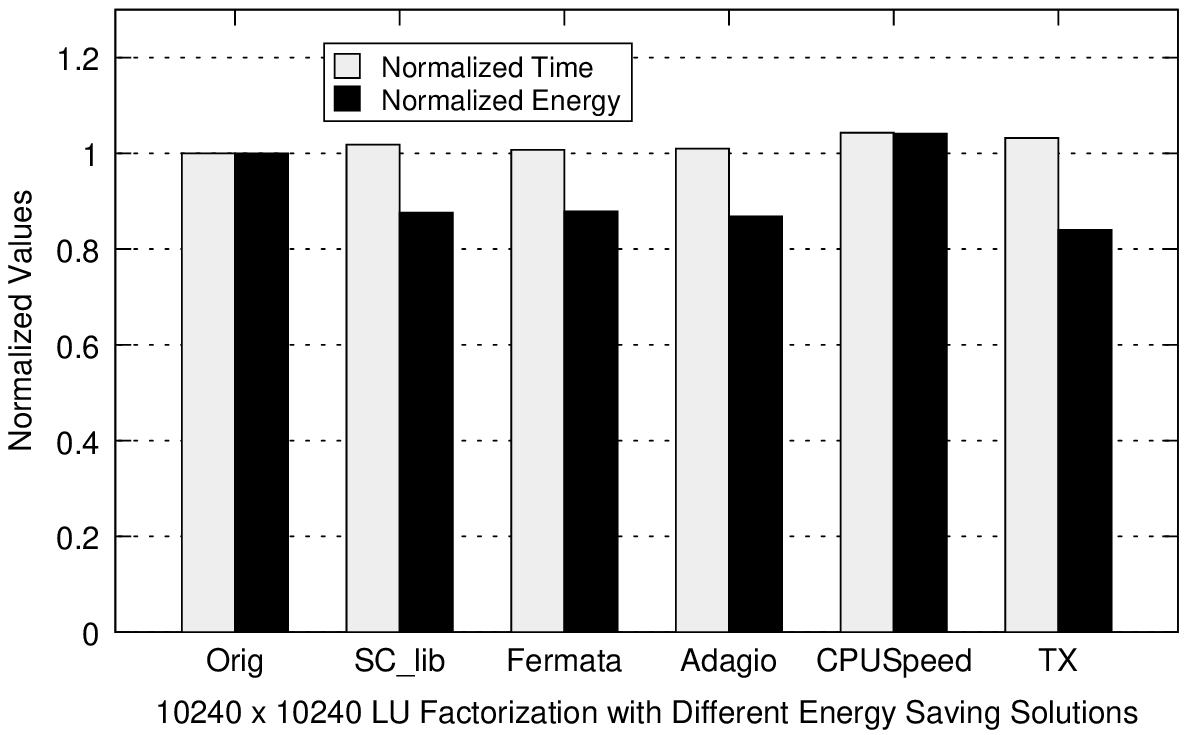} &
\hspace{-7mm}\includegraphics[width=.46\textwidth]{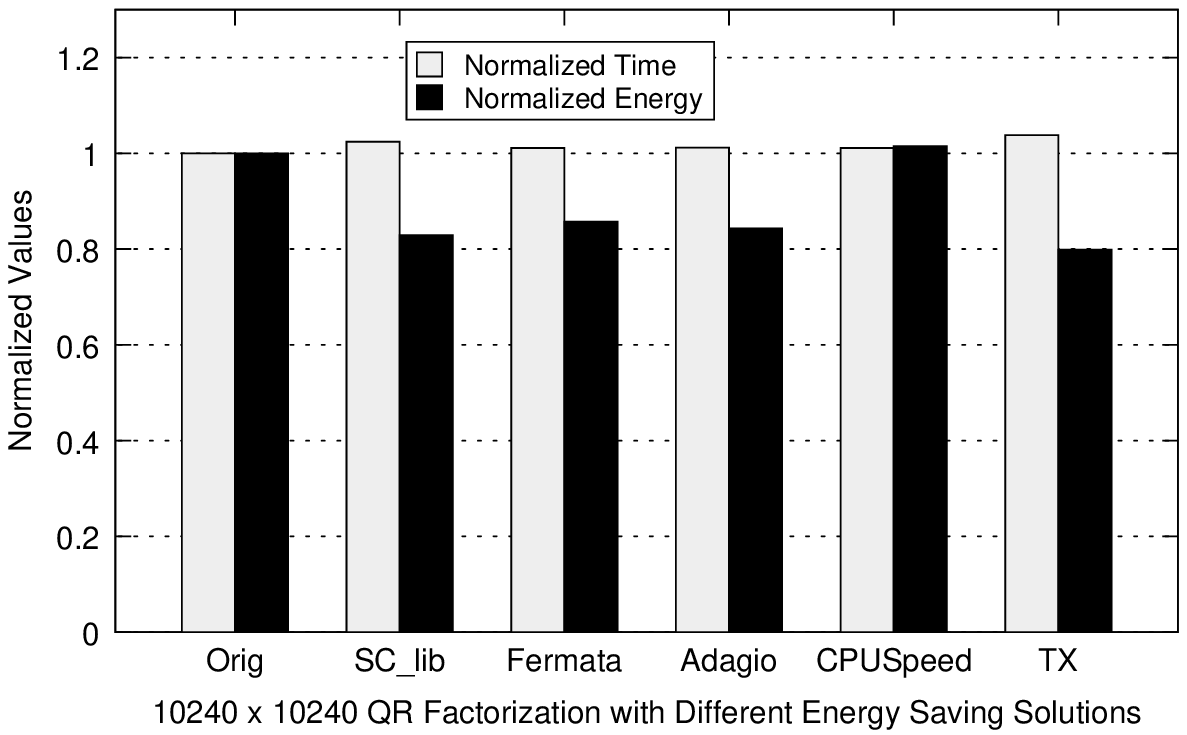}\\
\hspace{-22mm}\includegraphics[width=.46\textwidth]{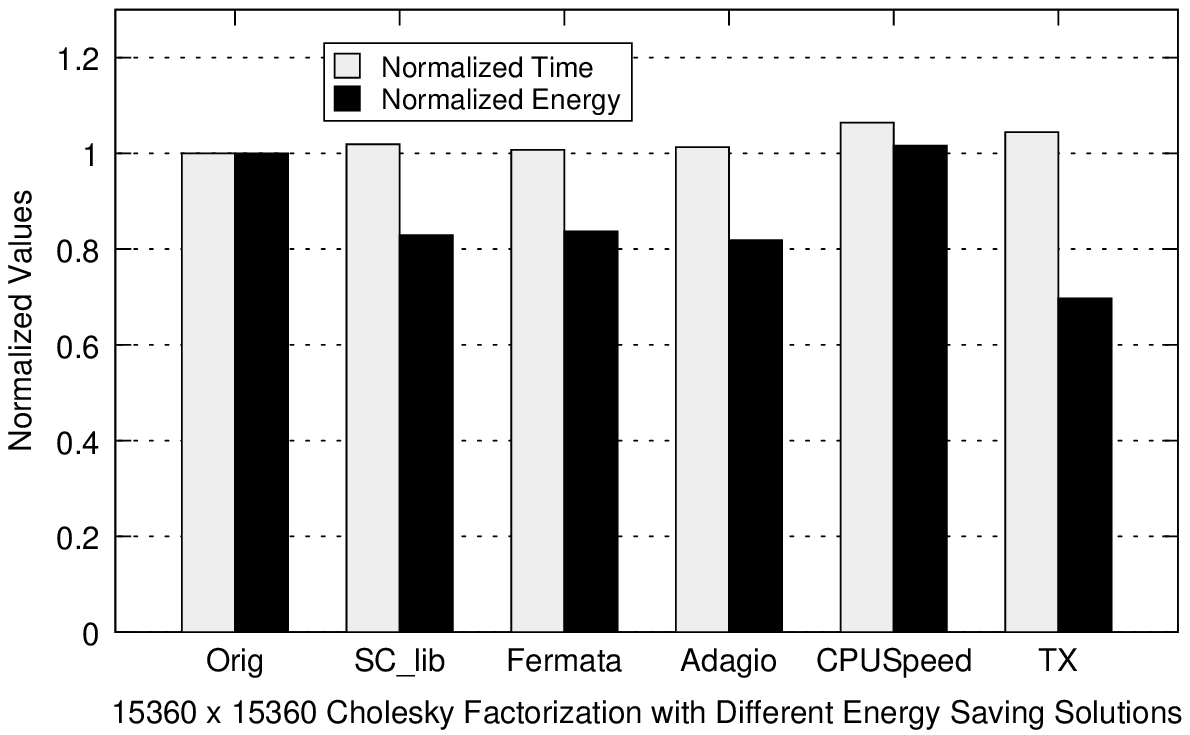} &
\hspace{-7mm}\includegraphics[width=.46\textwidth]{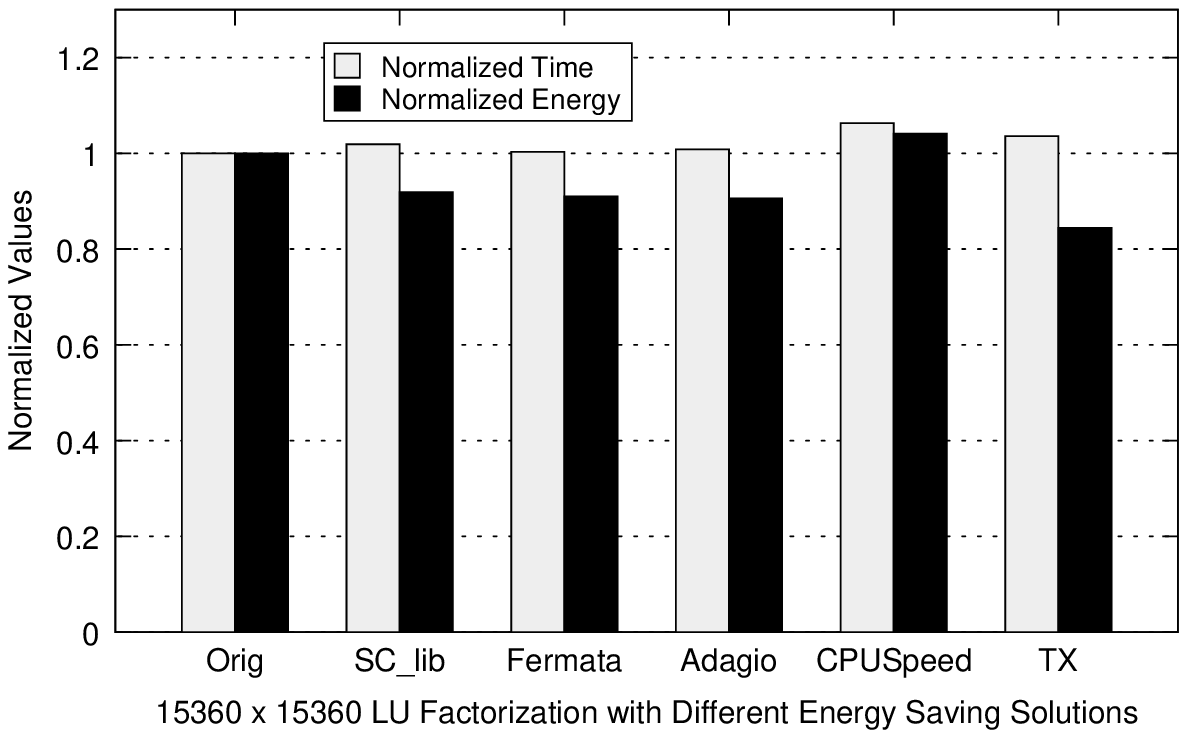} &
\hspace{-7mm}\includegraphics[width=.46\textwidth]{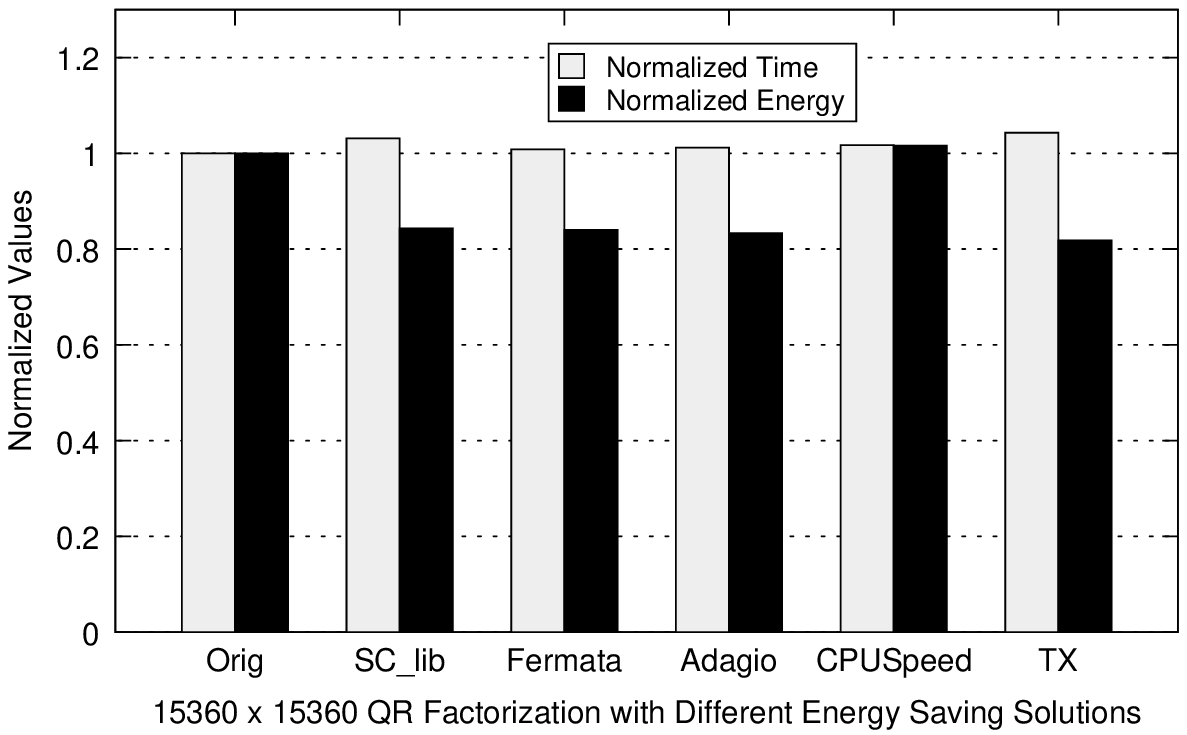}\\
\hspace{-22mm}\includegraphics[width=.46\textwidth]{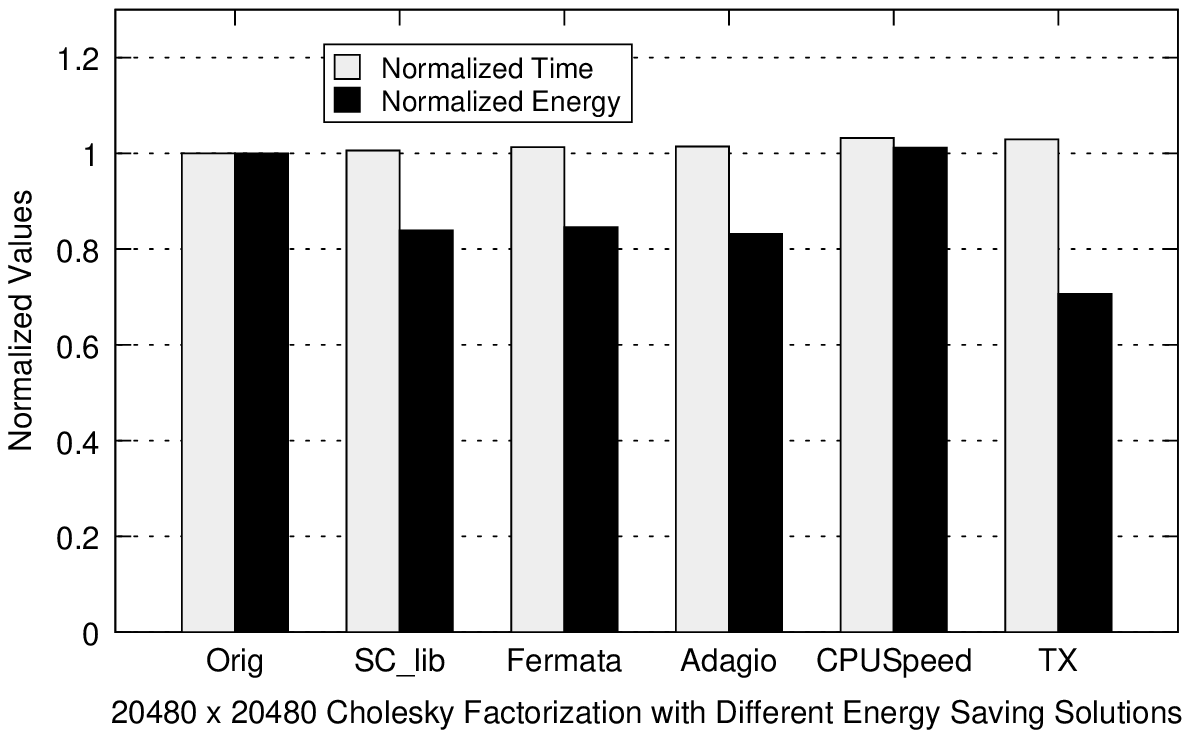} &
\hspace{-7mm}\includegraphics[width=.46\textwidth]{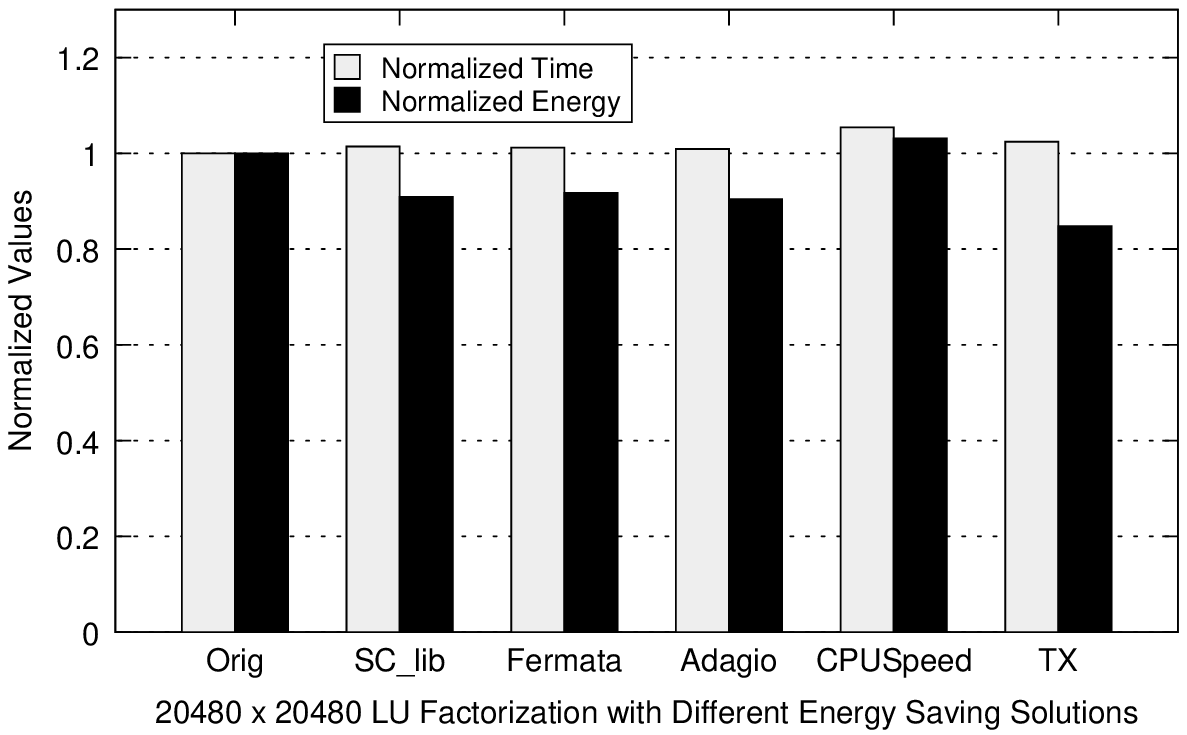} &
\hspace{-7mm}\includegraphics[width=.46\textwidth]{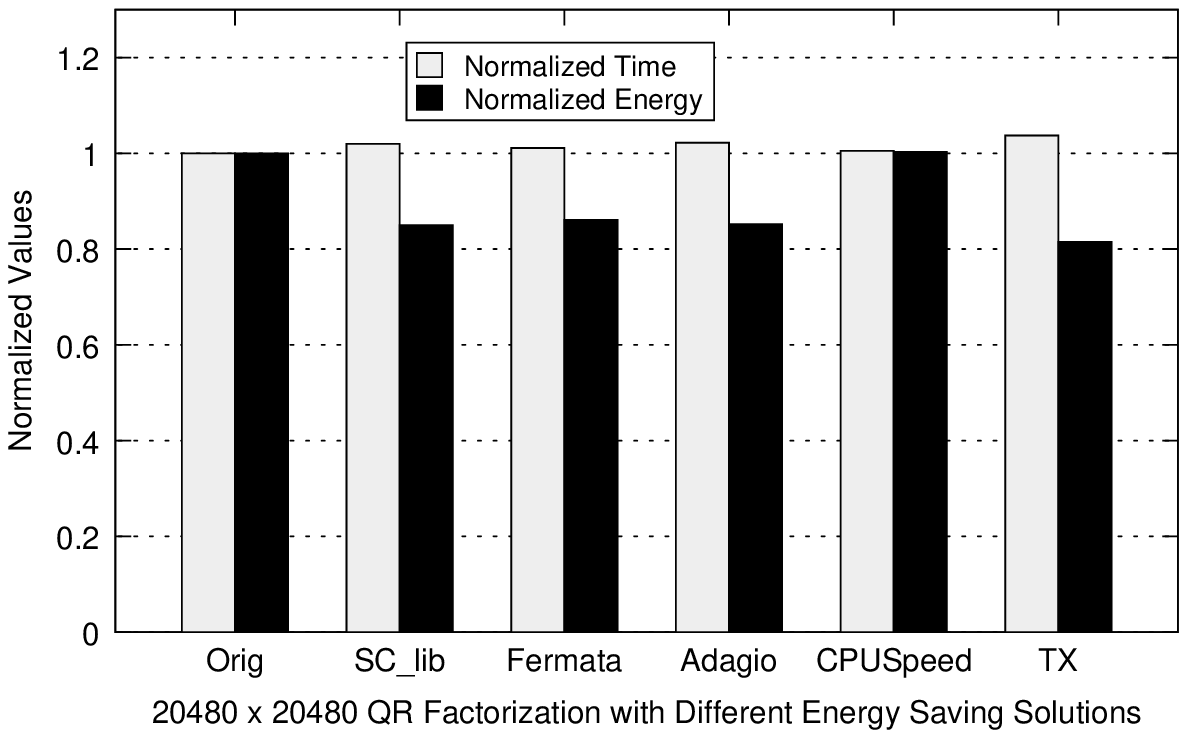}\\
\hspace{-22mm}\includegraphics[width=.46\textwidth]{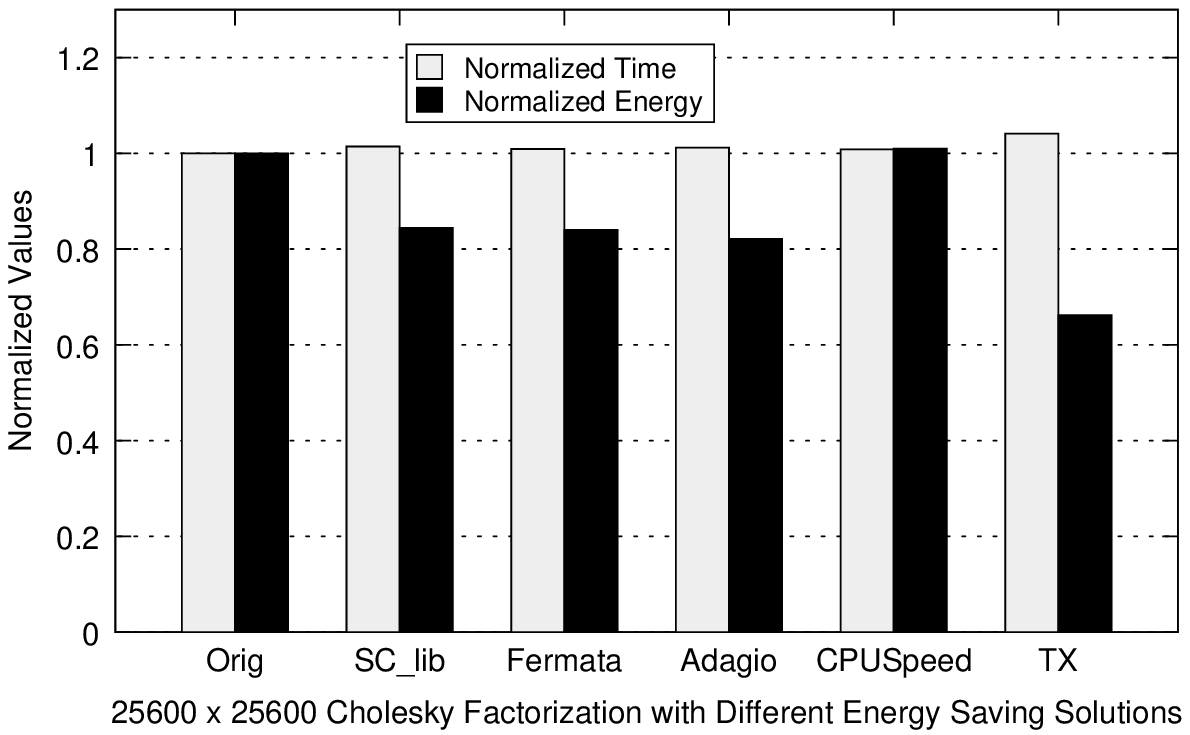} &
\hspace{-7mm}\includegraphics[width=.46\textwidth]{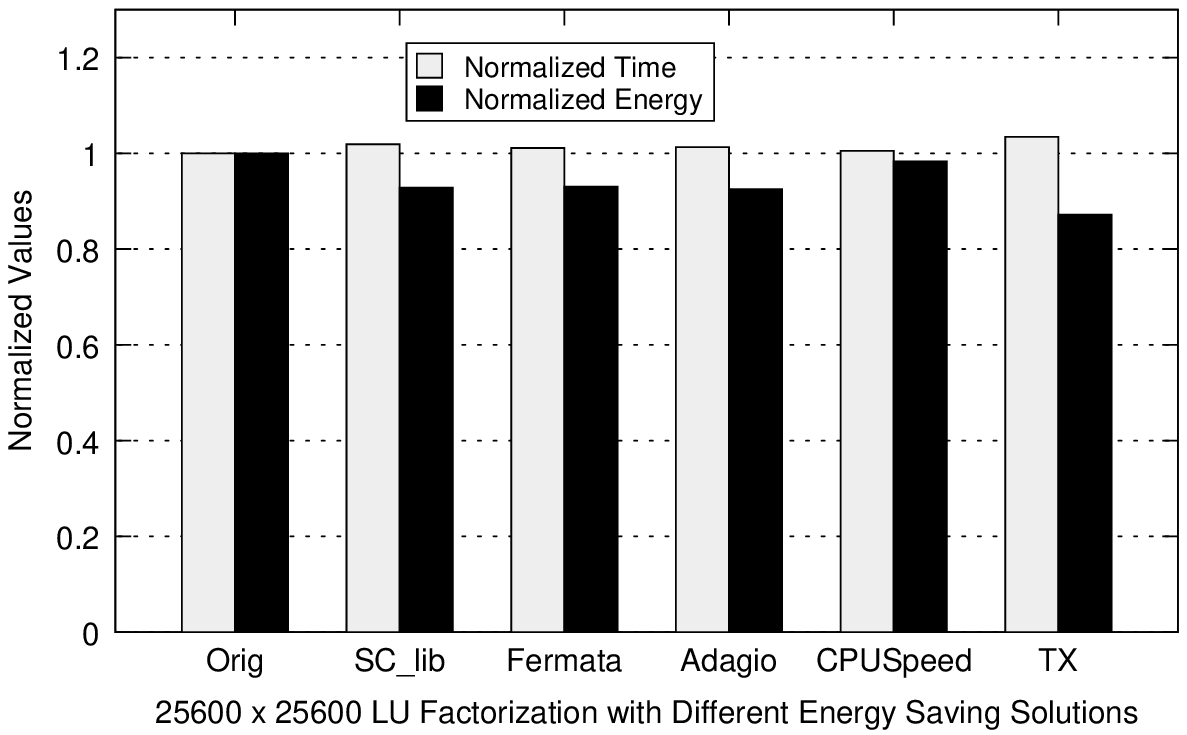} &
\hspace{-7mm}\includegraphics[width=.46\textwidth]{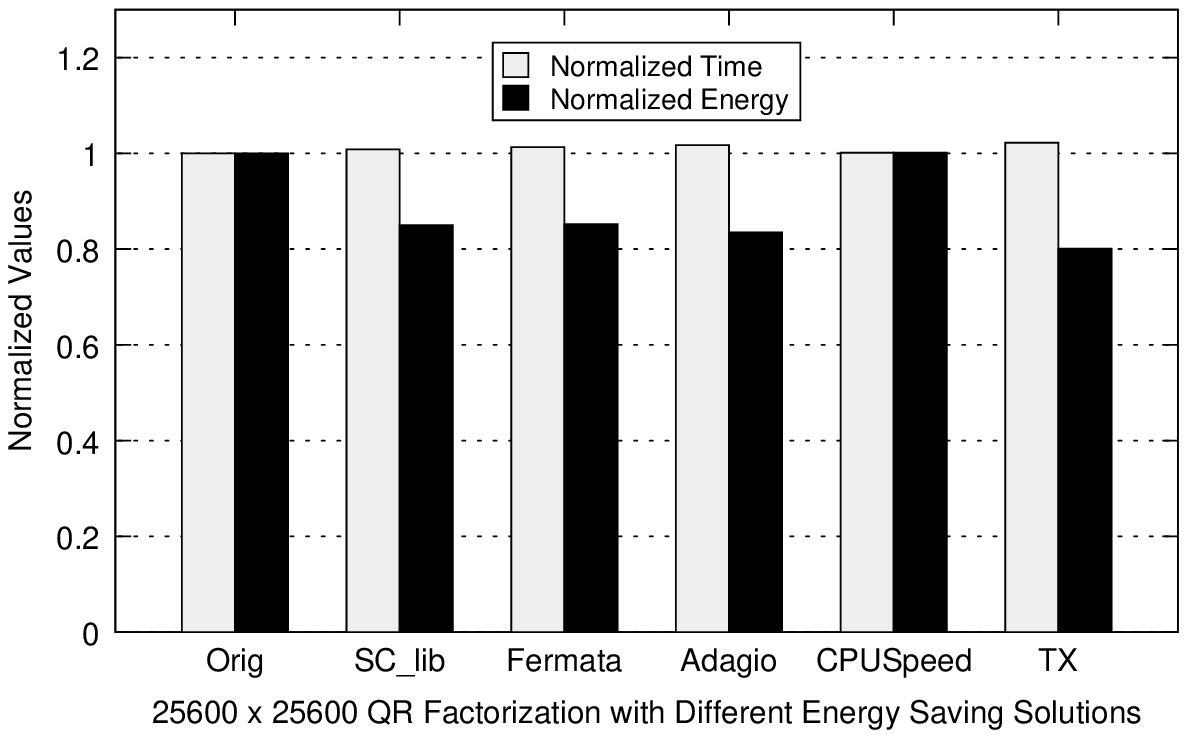}\\
\end{tabular}
\caption{Energy and Performance Efficiency of Parallel Cholesky, LU, and QR Factorizations on the HPCL Cluster with Different Global Matrix Sizes and Energy Saving Approaches using 8 $\times$ 8 Process Grid.}
\label{energy_performance_efficiency_hpcl}
\vspace{-0.0335mm}
\end{figure*}

\begin{figure*}
\vspace{-2mm}
\begin{tabular}{ccc}
\begin{minipage}{2.79in}
\centering
\hspace{-59mm}\includegraphics[width=2.79in]{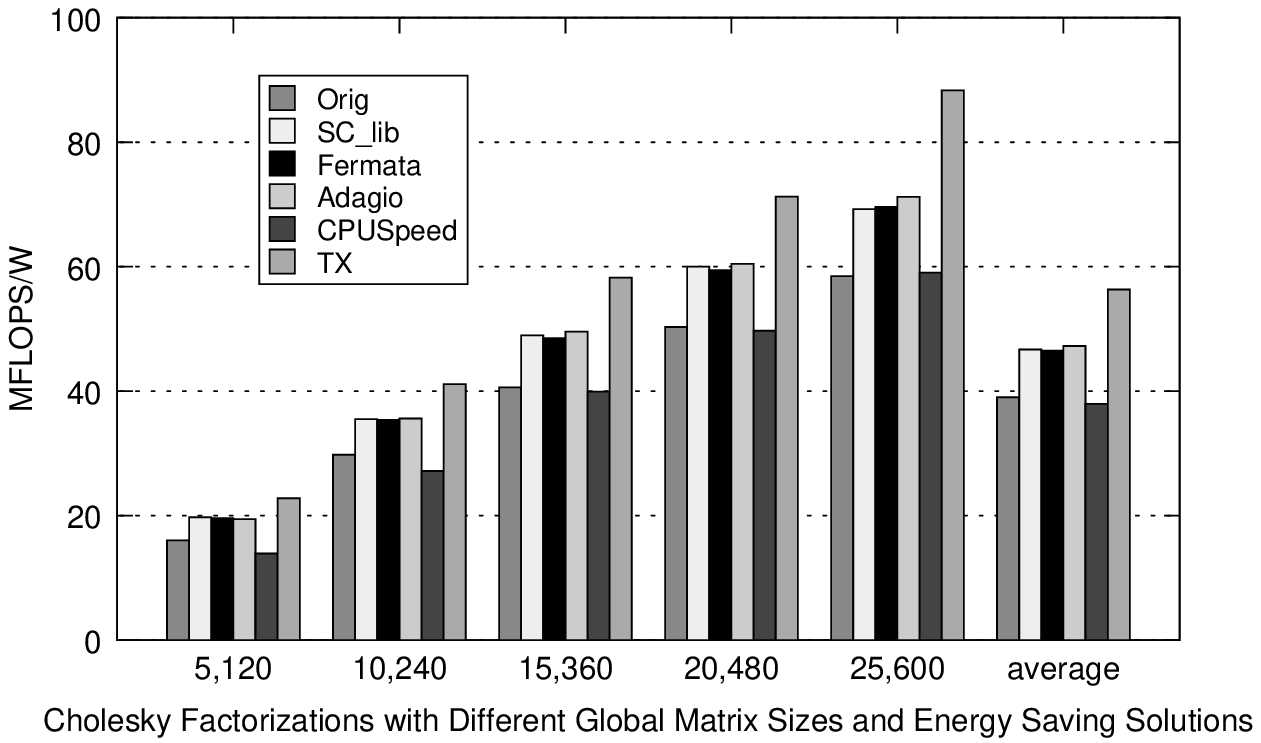}
\end{minipage}
&
\begin{minipage}{2.79in}
\centering
\hspace{-84mm}\includegraphics[width=2.79in]{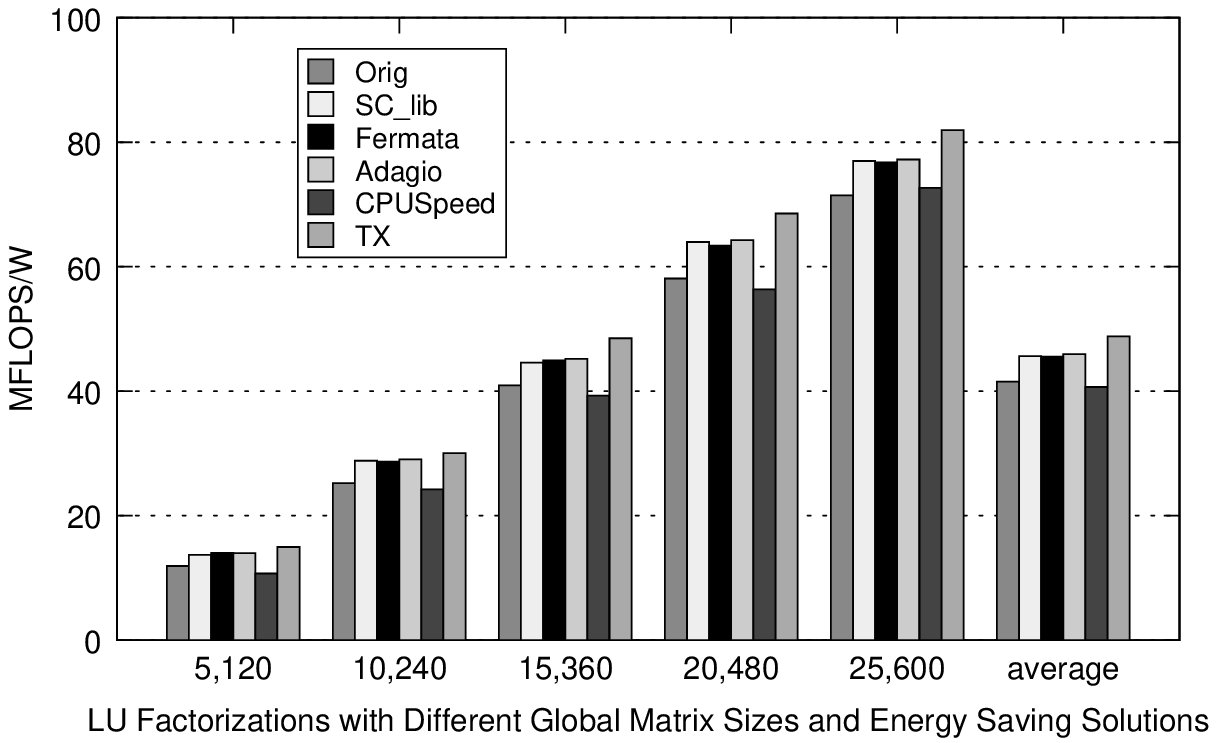}
\end{minipage}
&
\begin{minipage}{2.79in}
\centering
\hspace{-109mm}\includegraphics[width=2.79in]{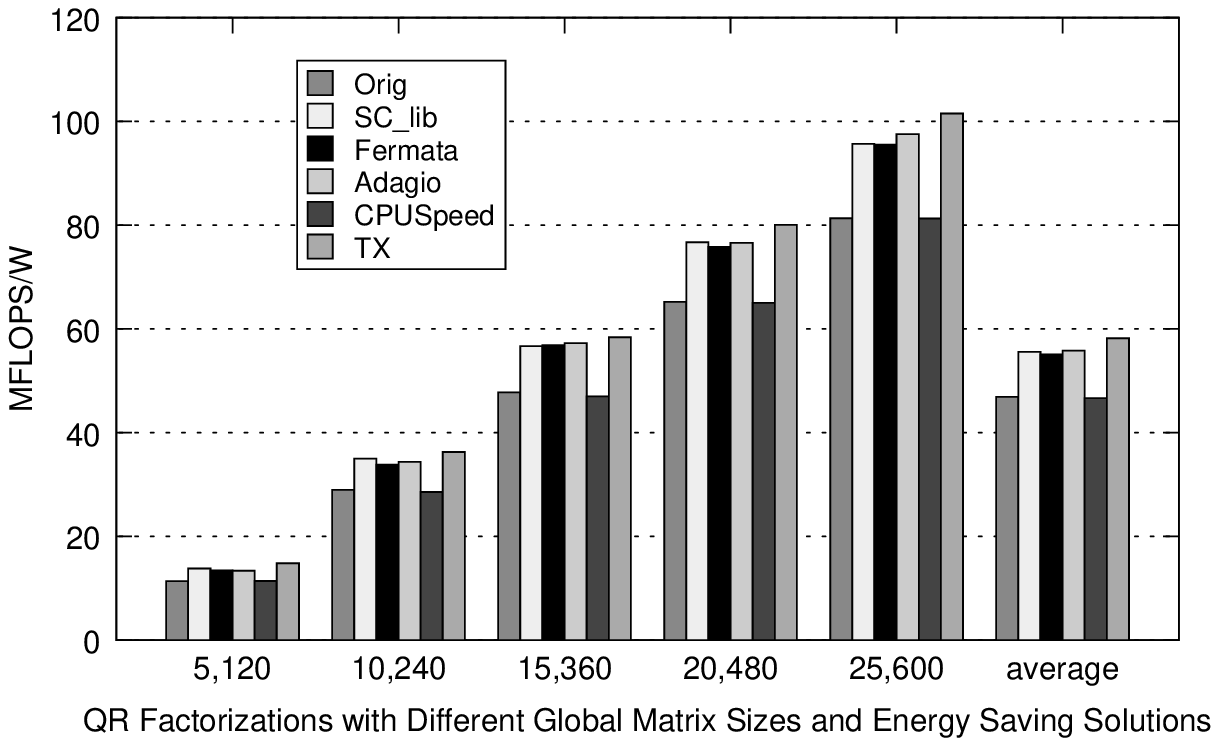}
\end{minipage}
\end{tabular}
\caption{Energy and Performance Trade-off of Parallel Cholesky, LU, and QR Factorizations on the HPCL Cluster with Different Global Matrix Sizes and Energy Saving Approaches using 8 $\times$ 8 Process Grid.}
\label{energy_performance_trade-off_hpcl}
\vspace{-3mm}
\end{figure*}

\vspace{1mm}
\noindent\textsc{\textbf{Energy Savings}}. Next we compare energy savings achieved by all five approaches (not including \texttt{CP\_theo}) on parallel Cholesky, LU, and QR factorizations on the HPCL cluster as shown in Figure \ref{energy_performance_efficiency_hpcl}, where the energy consumption was measured by recording on/off the collection of power and time costs when an application starts/ends. For eliminating errors from scalability, we collected energy and time data of five matrix factorizations with different global matrix sizes ranging from 5120 to 25600, respectively. Considerable energy savings are achieved by all approaches except for \texttt{CPUSpeed} on all Cholesky, LU, and QR with similar energy saving trend: \texttt{TX} prevails over all other approaches with higher energy efficiency, while \texttt{SC\_lib}, \texttt{Fermata}, and \texttt{Adagio} have similar energy efficiency. Overall, for Cholesky, \texttt{TX} can save energy 30.2\% on average and up to 33.8\%; for LU and QR, \texttt{TX} can achieve 16.0\% and 20.0\% on average and up to 20.4\% and 23.4\% energy savings, respectively. Due to the reasons discussed for power savings, \texttt{Adagio} only achieves similar energy savings as \texttt{SC\_lib} and \texttt{Fermata}, without fulfilling additional energy savings from slack reclamation of computation. With application-specific knowledge instead of workload prediction, \texttt{TX} manages to achieve energy savings during both computation and communication slack. Moreover, \texttt{TX} benefits from the advantage of saving additional energy during possible load imbalance other approaches cannot exploit. Next we further evaluate such energy savings by increasing load imbalance in the applications.

\vspace{1mm}
\noindent\textsc{\textbf{Effects of Block Size}}. As discussed earlier, the additional energy savings can be achieved from load imbalance, i.e., the durations only covered by green dashed boxes as shown in Figure \ref{Cholesky_DAG}. Empirically, regardless of the workload partition techniques employed, load imbalance can grow due to larger tasks, longer communication, etc. For manifesting the strength of \texttt{TX} in achieving additional energy savings for \emph{completeness}, we deliberately imbalance the workload through expanding tasks by using greater block sizes for Cholesky, while keeping the default block size for LU and QR. As shown in Figure \ref{energy_performance_efficiency_hpcl}, the average energy savings fulfilled by \texttt{TX} for Cholesky (30.2\%) are consequently greater than LU and QR (16.0\% and 20.0\%). Compared to the second most effective approach \texttt{Adagio}, \texttt{TX} can save Cholesky 12.8\% more energy on average.

\vspace{1mm}
\noindent\textsc{\textbf{Performance Loss}}. Figure \ref{energy_performance_efficiency_hpcl} also illustrates performance loss from different energy saving approaches against the original runs. We can see \texttt{TX} only incurs negligible time overhead: 3.8\%, 3.1\%, 3.7\% on average for Cholesky, LU, and QR individually, similar to the time overhead of all other approaches except for \texttt{CPUSpeed}. The minor performance loss on employing these solutions is primarily originated from three aspects: (a) Although large-message communication is not CPU-bound, pre-computation required for starting up a communication link before any data transmission is necessary and is affected by CPU performance, so the low-power state during communication can slightly degrade performance; (b) switching CPU frequency via DVFS is essentially implemented by modifying CPU frequency system configuration files, and thus slight time overhead is incurred from the in-memory file read/write operations \cite{iccs14}; and (c) CPU frequency transition latency is required for the newly-set frequency to take effect. Further, \texttt{TX} suffers from minor performance loss from TDS analysis, including TDS generation and maintaining TDS for each task. The high time overhead of \texttt{CPUSpeed} is another reason for its little and even negative energy savings besides the defective prediction mechanism at OS level.

\vspace{1mm}
\noindent\textsc{\textbf{Energy/Performance Trade-off}}. An optimal energy saving approach requires to achieve the maximal energy savings with the minimal performance loss. Per this requirement, energy-performance integrated metrics are widely employed to quantify if the energy efficiency achieved and the performance loss incurred meanwhile are well-balanced. We adopt Energy-Delay Product (EDP) to evaluate the overall energy and performance trade-off of the five approaches, in terms of MFLOPS/W, which equals the amount of floating-point operations per second within the unit of one Watt, i.e., the greater value it is, the better efficiency is fulfilled. As shown in Figure \ref{energy_performance_trade-off_hpcl}, compared to other approaches, \texttt{TX} is able to fulfill the most balanced trade-off between the energy and performance efficiency achieved. Specifically, \texttt{TX} has higher MFLOPS/W values for Cholesky compared to LU and QR, due to the higher energy savings achieved from the more imbalanced workload in Cholesky without additional performance loss.

\section{Related Work}

Numerous other types of energy efficient DVFS scheduling algorithms have been proposed, but only a few of them were designed for high performance scientific computing. We next detail them in the following categories \emph{OS-level}, \emph{Application-level}, \emph{Simulation-based}.

\vspace{1mm}
\noindent\textsc{\textbf{OS-level}}. There exist a large body of OS level energy efficient approaches for high performance scientific applications. Lim \textit{et al.} \cite{sc06} developed a runtime system that dynamically and transparently reduces CPU power for communication phases to minimize energy-delay product. Ge \textit{et al.} \cite{icpp07} proposed a runtime system and an integrated performance model for achieving energy efficiency and constraining performance loss through performance modeling and prediction. Rountree \textit{et al.} \cite{sc07} developed a \emph{SC} approach that employs a linear programming solver collecting communication trace and power characteristics for generating an energy saving scheduling. Subsequent work \cite{ics09} presented another runtime system by improving and extending previous classic scheduling algorithms and achieved significant energy savings with extremely limited performance loss.

\vspace{1mm}
\noindent\textsc{\textbf{Application-level}}. Kappiah \textit{et al.} \cite{sc05b} introduced a scheduled iteration method that computes the total slack per processor per timestep, then scheduling CPU frequency for the upcoming timestep. Liu \textit{et al.} \cite{ipdps05b} presented a technique that tracks the idle durations for one processor to wait for others to reach the same program point, and utilizes this information to reduce the idle time via DVFS without performance loss. Tan \textit{et al.} \cite{ipccc13} proposed an adaptively aggressive scheduling strategy for data intensive applications with moderated performance trade-off using speculation. Subsequent work \cite{iccs14} proposed an adaptive memory-aware strategy for distributed matrix multiplication that trades grouped computation/communication with memory costs for less overhead on employing DVFS. Liu \textit{et al.} \cite{pact12} proposed a power-aware static mapping technique to assign applications for a CPU/GPU heterogeneous system that reduces power and energy costs via DVFS on both CPU and GPU, with timing requirements satisfied.

\vspace{1mm}
\noindent\textsc{\textbf{Simulation-based}}. There exist some efforts on improving energy efficiency for numerical linear algebra operations like Cholesky/LU/QR factorization, but most of them either are based on simulation or only work for a single multicore machine. Few studies have been conducted on power/energy efficient matrix factorizations running on distributed-memory architectures. Slack reclamation methods such as Slack Reduction and Race-to-Idle algorithms \cite{csrd12a} \cite{hpcs11} have been proposed to save energy for dense linear algebra operations on shared-memory multicore processors. Instead of running benchmarks on real machines, a power-aware simulator, in charge of runtime scheduling to achieve task level parallelism, was employed to evaluate the proposed power-control policies for linear algebra operations. DVFS techniques used in their approaches were also simulated. Subsequent work \cite{pdp12a} leveraged DVFS to optimize task-parallel execution of a collection of dense linear algebra tasks on shared-memory multicore architectures, where experiments were performed at thread level. Since no communication was involved, these approaches did not achieve significant energy savings due to no utilization of slack from communication latency.

\section{Conclusions}

The looming overloaded energy consumption of high performance scientific computing brings significant challenges to green computing in this era of ever-growing power costs for large-scale HPC systems. DVFS techniques have been widely employed to improve energy efficiency for task-parallel applications. With high generality, OS level solutions are regarded as feasible energy saving approaches for such applications. We observe for applications with variable execution characteristics such as parallel Cholesky, LU, and QR factorizations, OS level solutions suffer from the defective prediction mechanism and untapped potential energy savings from possible load imbalance, and thus cannot optimize the energy efficiency. Giving up partial generality, the proposed library level approach \texttt{TX} is evaluated to save more energy with little performance loss for parallel Cholesky, LU, and QR factorizations on two power-aware clusters compared to classic OS level solutions.




\bibliographystyle{elsarticle-num}
\bibliography{elsarticle}







\end{document}